\documentclass[usenatbib]{mnras}

\usepackage[T1]{fontenc}
\usepackage{ae,aecompl,color}

\usepackage{graphicx}      
\usepackage{epstopdf}
\usepackage{amsmath}    
\usepackage{amssymb}    
\usepackage{booktabs}
\usepackage{bm}

\definecolor{green}{rgb}{0,0.5,0.0}
\definecolor{navy}{rgb}{0.0,0.0,0.6}
\definecolor{orange}{rgb}{0.8,0.2,0.0}

\usepackage[normalem]{ulem}  

\def\apjl{ApJ Lett.}%
\def\aap{A\&A}%
\def\mnras{MNRAS}%
\def\prc{Phys. Rev. C\ }%
\def\prd{Phys. Rev. D \ }%
\def\nat{Nature}%
\title[Cooling of hypernuclear stars ]{                                                                                                   
Cooling of hypernuclear compact stars: Hartree-Fock models and high-density pairing}

\author[A. R. Raduta et al. ]{
Adriana R. Raduta$^{1}$\thanks{E-mail: araduta@nipne.ro}
Jia Jie Li$^{2}$\thanks{E-mail: jiajieli@th.physik.uni-frankfurt.de}
Armen Sedrakian $^{3,4}$\thanks{sedrakian@fias.uni-frankfurt.de}
and Fridolin Weber $^{5,6}$\thanks{fweber@sdsu.edu}
\\
$^{1}$National Institute for Physics and Nuclear Engineering,
RO-077125 Bucharest, Romania\\
$^{2}$Institute for Theoretical Physics, J. W. Goethe University, D-60438 \
Frankfurt am Main, Germany \\
$^{3}$Frankfurt Institute for Advanced Studies, D-60438
Frankfurt-Main, Germany\\
$^{4}$Institute of Theoretical Physics, University of Wroclaw,
  50-2014 Wroclaw, Poland\\
$^{5}$Department of Physics, San Diego State University, 5500
Campanile Drive, San Diego, CA 92182, USA \\
$^{6}$Center for Astrophysics and Space Sciences, University of California at San
Diego, La Jolla, CA 92093, USA 
}

\begin{document}
\label{firstpage}
\pagerange{\pageref{firstpage}--\pageref{lastpage}}
\maketitle

\begin{abstract}
  The thermal evolution of hypernuclear compact stars is studied for
  stellar models constructed on the basis of covariant density
  functional theory in Hartree and Hartree-Fock approximation. 
  Parametrizations of both types are consistent with the astrophysical mass
  constraints on compact stars and available hypernuclear data. We
  discuss the differences of these density functionals and highlight
  the effects they have on the composition and on the cooling of
  hypernuclear stars.  It is shown that hypernuclear stars computed
  with density functional models that have a low symmetry energy
  slope, $L$, are fairly consistent with the cooling data of observed
  compact stars.  The class of stellar models based on larger $L$
  values gives rise to the direct Urca process at low densities, which
  leads to significantly faster cooling. We conjecture high-density
  pairing for protons and $\Lambda$'s in the $P$-wave channel and
  provide simple scaling arguments to obtain these gaps.  As a
  consequence the most massive stellar models with masses
  $1.8 \le M/M_{\odot} \le2$ experience slower cooling by hyperonic
  dUrca processes which involve $\Lambda$'s and protons.
\end{abstract}


\section{Introduction}

The interest in the hyperonization of matter at high densities in
compact stars was rekindled in the last decade by the observations of
several white dwarf--pulsar binaries with pulsar masses close to two
solar masses
\citep{2010Natur.467.1081D,2013Sci...340..448A,Fonseca2016,2017MNRAS.465.1711B}.
This problem was studied mainly within the density functional theory
based on the Hartree (mean-field) approximation to obtain the equation
of state (EoS) of
matter~\citep{Weissenborn2012b,Weissenborn2012a,Bonanno2012,Bednarek2012,Long2012,Colucci2013,Miyatsu_PRC2013,Dalen2014,Gusakov_MNRAS2014,Maslov2015,Gomes2015Ap,Oertel2015,Fortin_PRC2016,Tolos2016,Fortin2017,Marques2017,Spinella-PhD,Li2018EPJA,Li2018PLB,SpinellaWeber2018}.
Two solar mass neutron star models with a hyperonic admixture
  have nevertheless been constructed also within alternative frameworks,
  which include the quark-meson coupling model~\citep{Stone_NPA_2007},
  the auxiliary field diffusion Monte-Carlo
  approach~\citep{Lonardoni_PRL_2015}, cluster variational method
  \citep{Togashi_PRC_2016,Yamamoto_PRC_2017} and, more recently,
  Brueckner-Hartree-Fock theory \citep{Yamamoto_PRC_2017}.   Density
functionals for hypernuclear matter and its EoS were recently
constructed on the basis of Hartree-Fock
theory~\citep{Li2018EPJA,Li2018PLB}.  Including the Fock contribution
in the treatment allows one to introduce the pion (and more generally
tensor forces) explicitly, thus achieving a more versatile
representation of the density functional.  In this work we will
include these EoS in our study of the cooling behavior of hypernuclear
stars thus extending the pool of EoS from which our models are
built~\citep{Raduta2018MNRAS}; hereafter Paper I.  Another purpose of
this work is to examine how changes in the patterns of the
high-density pairing affect the cooling of hypernuclear stars. It was
observed that the most massive hypernuclear stars cool very fast by
the $p\Lambda$ direct Urca (hereafter dUrca) process (see Paper I).
The fast cooling arises because these stars develop cores where
protons and $\Lambda$'s are unpaired, as their partial densities
exceed the density at which the $^1S_0$ pairing gap closes. As is well
known, neutron matter supports $^3P_2$-$^3F_2$ pairing at densities
beyond the density for $^1S_0$ pairing. We conjecture here that the
same occurs for the protonic and $\Lambda$ components of dense
matter. Below, we will use simple estimates of the gaps for protons
and $\Lambda$'s, which are based on the symmetries of nuclear and
hypernuclear forces, to gauge the effect of high-density pairing on
the cooling of massive hypernuclear stars.

The early studies of the cooling of hypernuclear stars did not include
the high-mass astrophysical constraint on the EoS of hypernuclear
matter \citep{Haensel_1994,Schaab_1998,Tsuruta_2009}.  In Paper I we
started a systematic study of the cooling of hypernuclear compact
stars on the basis of modern density functionals mentioned
above. Several works have appeared since then in this context:
\cite{Grigorian_NPA_2018} used the ``nuclear medium cooling''
scenario~\citep{Blaschke2004,Schaab1997AA} to account for hyperons
using an EoS based on the density functional
theory~\citep{Maslov2015}, pion-mediated enhanced neutrino
emissivities for the modified Urca and bremsstrahlung processes,
vanishing $P$-wave neutron pairing, and phenomenological crust-core
temperature relation.  The pion-mediated enhanced neutrino
emissivities and vanishing $P$-wave neutron pairing are the key
factors that distinguish their models from those in Paper I.
Nevertheless, the remainder of the physical input, for example, the
crust-core temperature relation, is not identical to ours. Note that
Paper I does not implement the minimal cooling
scenario~\citep{Page_2004} as it allows for the direct dUrca process
among nucleons and hyperons, uses pair-breaking processes with dressed
(in-medium) vertices, and covers a wide range of masses up to the
maximum mass of $M_{\rm max}/M_{\odot}\sim 2$, which implies a varying
cooling behavior for the mass hierarchy (see, in particular, the
conclusion Section of Paper I). The same problem was also addressed by
\cite{Tolos2018ApJ} using hyperonic EoS based on the FSU class of
relativistic density functional models which feature massive objects
and are tuned to available data on
hypernuclei~\citep{Tolos2017a,Tolos2017b}.  In these models, the main
cooling agents are the ($\Lambda,p$) and ($n,p$) dUrca processes which
were regulated by variations of proton and neutron superfluid gaps. In
particular the density range over which the proton $S$-wave gap is
non-zero has been explored with the aim to obtain a satisfactory
agreement with the data.  The main difference with respect to Paper I
is the absence of pairing in the hyperonic sector.  Combined, the
emerging new generation of cooling models of hypernuclear stars have
the potential to further constraining the physics of dense nuclear
matter, especially its {\it composition}.

This work is structured as follows. In Sec.~\ref{sec:EoS} we review
the EoS models used in our cooling simulations. We discuss the stellar
parameters (mass, radius, etc.) and the range of Fermi momenta of
dominant particles predicted by these models.
Section~\ref{sec:pairing} is devoted to the discussion of $P$-wave
pairing of protons and $\Lambda$'s. Simple symmetry arguments are used
to extract the values and density dependences of the corresponding
gaps. The critical temperatures of various species are mapped on the
internal radius of the models and the pairing patterns inside the star
{ are} discussed.  The results of numerical simulations of the thermal
evolution of hypernuclear compact stars are shown in
Section~\ref{sec:therm_ev}.  A short summary of our results along with
conclusions is given in Section~\ref{sec:conclusions}.

\section{Equation of state and pairing}
\label{sec:EoS}

\begin{figure}
\begin{center}
\includegraphics[angle=0, width=0.99\columnwidth]{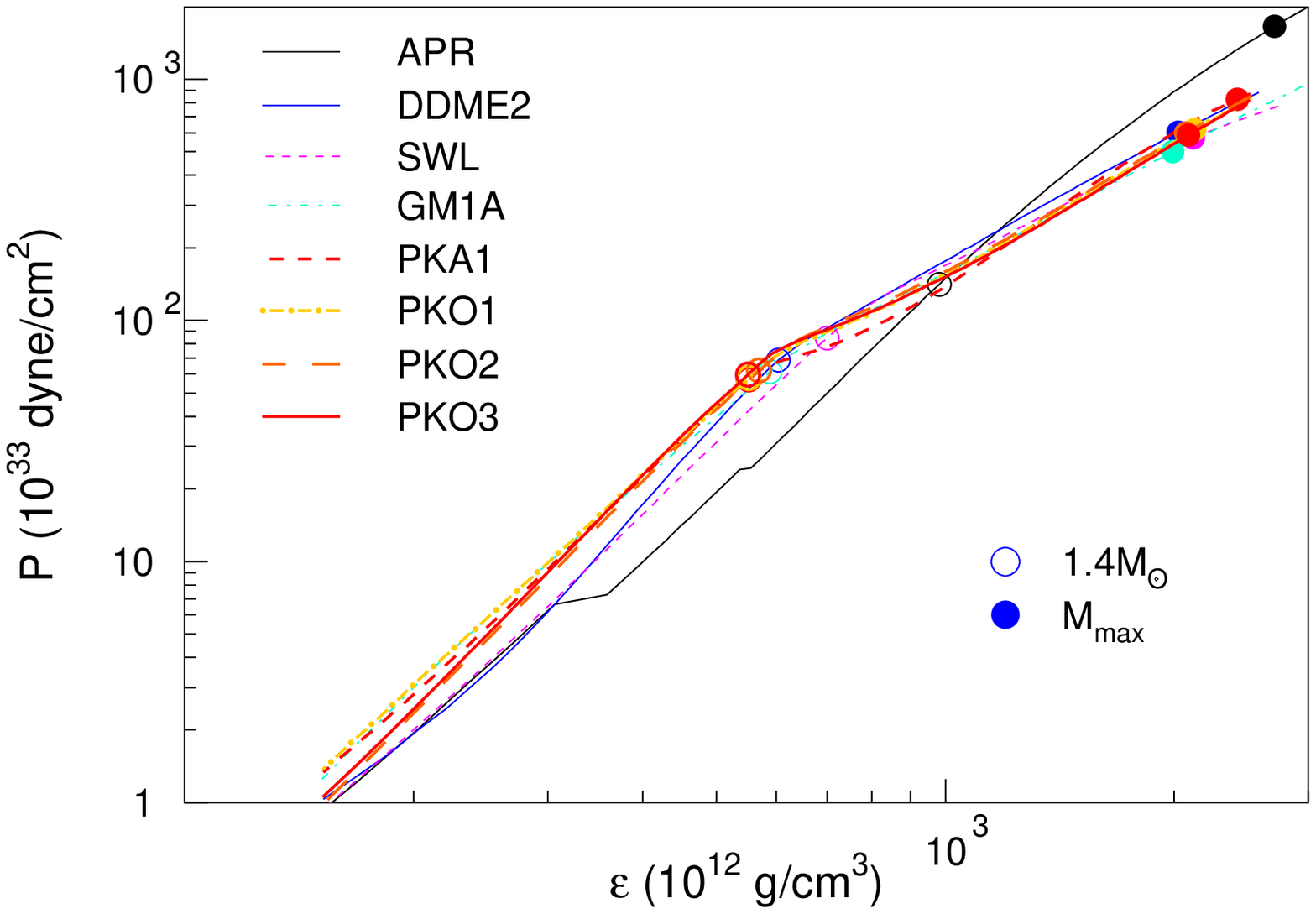}
\includegraphics[angle=0, width=0.99\columnwidth]{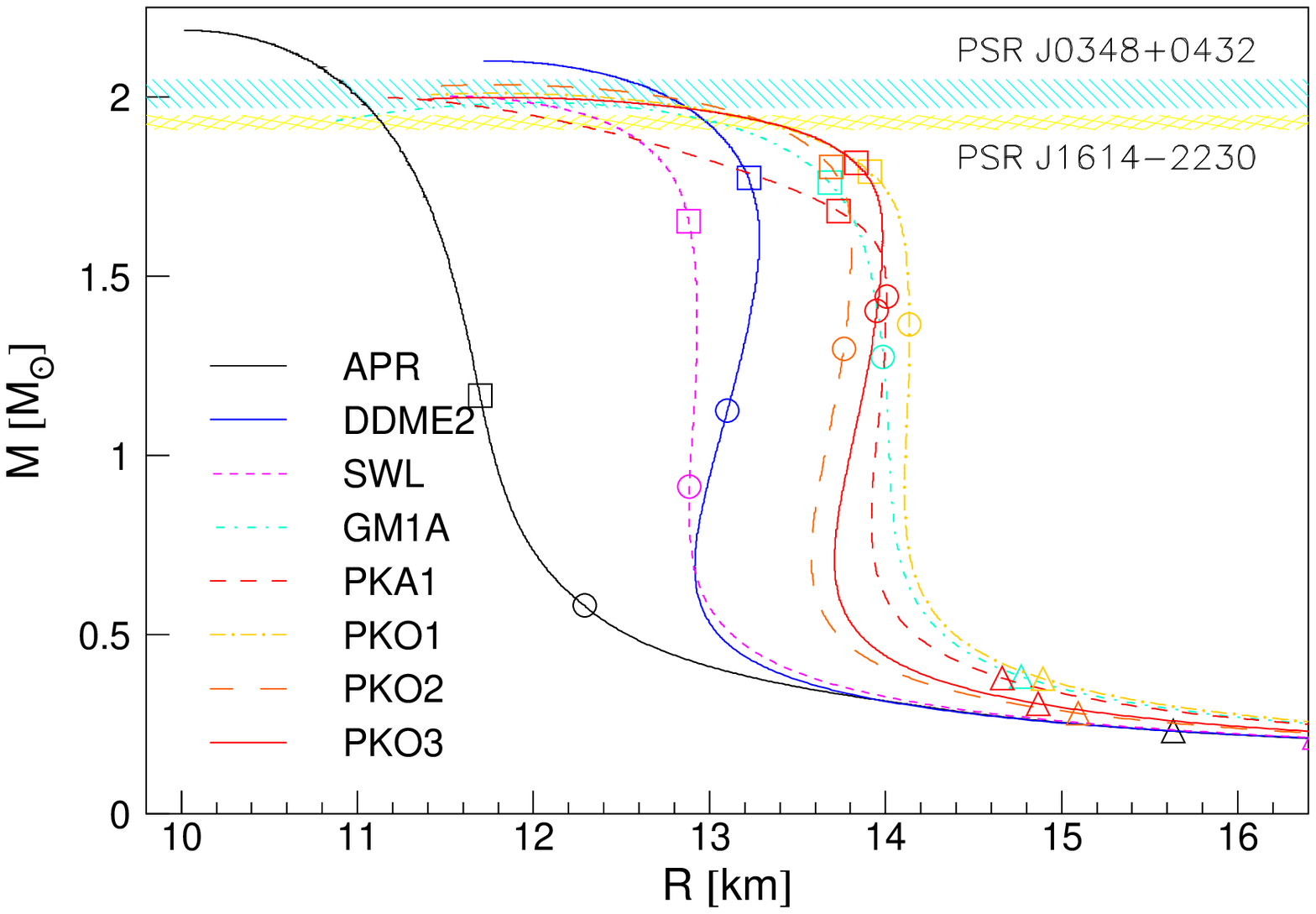}
\includegraphics[angle=0, width=0.99\columnwidth]{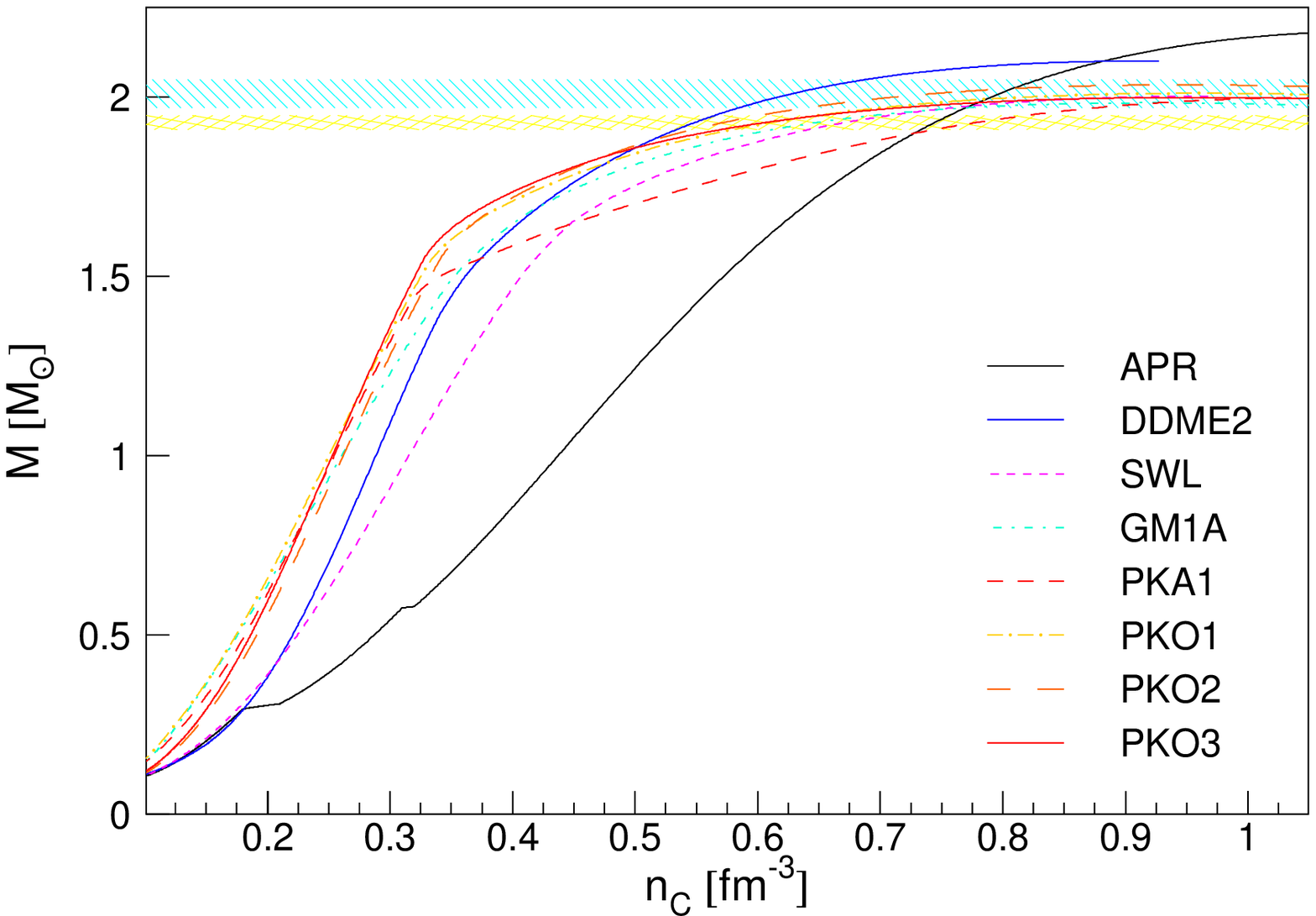}
\end{center}
\caption{Equation of state (top), mass vs. radius relation (middle),
  and mass vs. central baryonic number density (bottom) for the
  stellar models considered in this work.  For comparison,  the
  predictions of the purely nucleonic APR EoS \citep{APR} are
  represented as well. The solid (open) dots in the $P(\epsilon)$ plot
  show central values in maximum-mass ($1.4\, M_{\odot}$) neutron stars (NSs).
  In the $M-R$ diagram open
  triangles, dots and squares correspond to NS whose central densities
  equal $n_0$, $2 n_0$, and $3 n_0$, respectively.  The shaded
  horizontal stripes in the middle and bottom panels show the observed
  masses of pulsars PSR J0348+0432 ($M=2.01\pm 0.04 M_{\odot}$
  \citep{2013Sci...340..448A}) and PSR J1614-2230
  ($M=1.93\pm 0.02 M_{\odot}$ \citep{Fonseca2016}).  }
\label{fig:EOS}
\end{figure}

We consider a set of equations of state which describe hypernuclear
matter within the relativistic density functional theory (DFT) of
nuclear matter. In addition to the three Hartree models that were
studied in Paper I, here we add to our collection four Hartree-Fock
(HF) models for the EoS of hypernuclear
matter~\citep{Li2018EPJA,Li2018PLB}.  In contrast to Hartree models,
the HF models account for the tensor interaction between the baryons
and include $\pi$-meson exchanges in addition to the $\sigma$,
$\omega$, $\rho$ and possibly $\phi$ meson exchange included in the
Hartree models.  The key nuclear properties of these models are
displayed in Table~\ref{tab:models}. With the nuclear saturation
density ($n_0$) and energy per particle ($E_s$) well constrained, the list shows the
range of values of the nuclear compressibility $K$, symmetry energy
$J$, and its slope $L$. Note that all HF models are characterized by
large values of $L$ with the PKA1 parametrization having the largest
$J$ and $L$ values. These are at the boundaries inferred from 
experimental nuclear physics data \citep{Tsang_2012,Lattimer_2014}. 
Table~\ref{tab:rmfY} lists the
meson content of each model.  The vector meson-hyperon coupling
constants were chosen in each case on the basis of either the SU$(3)$
or SU$(6)$ spin-flavor symmetric quark model (see column 3 of
Table~\ref{tab:rmfY}). The coupling of the scalar meson to hyperons
are {tuned} such as to reproduce the depths of the hyperon potential
in symmetric saturated nuclear matter.  DDME2, GM1A and SWL use the
values $U_{\Lambda}^{(N)} \approx -28$ MeV,
$U_{\Xi}^{(N)} \approx -18$ MeV and $U_{\Sigma}^{(N)} \approx 30$ MeV
\citep{Gal_RMP_2016}, while { the PK parametrizations} use
$U_{\Lambda}^{(N)} \approx -30$ MeV, $U_{\Xi}^{(N)} \approx -14$ MeV
and $U_{\Sigma}^{(N)} \approx 30$ MeV. We note here the
sensitivity of our results to the value of the $\Sigma$ hyperon
potential in symmetric saturated nuclear matter. 
Changing it from $U_{\Sigma}^{(N)} \simeq 30$~MeV 
to the less repulsive value of $U_{\Sigma}^{(N)} \simeq 10$~MeV
lowers, within DDME2, the $\Sigma^-$ threshold below that for the $\Lambda$
hyperon~\citep{Colucci2013,Constanca_USN}.

\begin{table}
\begin{tabular}{lcccccc}
\hline 
Model& $n_{0}$       & $E_{s}$ & $K$  & $J$  & $L$  
                   \\
     & [fm$^{-3}$] & [MeV] & [MeV] & [MeV] & [MeV] \\
                   \hline
DDME2 & 0.152 & $-16.1$ & 250.9 & 32.3 &  51.2      \\
SWL   & 0.150 & $-16.0$ & 260.0 & 31.0 & 55.0     \\
GM1A  & 0.154 & $-16.3$ & 300.7 & 32.5 & 94.4  \\
PKA1  & 0.160 & $-15.8$ & 230.0 & 36.0 & 103.5 \\
PKO1  & 0.152 & $-16.0$ & 250.3 & 34.4 &  97.7 \\
PKO2  & 0.151 & $-16.0$ & 249.5 & 32.5 &  75.9  \\
PKO3  & 0.153 & $-16.0$ & 262.4 & 33.0 &  83.0 \\
APR & 0.160 & $-16.0$ & 266.0 & 32.6 & 57.6\\
\hline                                                                                                                                    
\end{tabular}
\caption{Key nuclear matter properties of the relativistic DF models
  considered in this work.
  DDME2, SWL, and GM1A are relativistic Hartree models,
  PKA1, PKO1,  PKO2,  and PKO3 are Hartree-Fock models. 
  Listed are the energy per nucleon ($E_s$)
  and compression modulus ($K$) at the saturation density of symmetric
  nuclear matter ($n_0$) together with the symmetry energy ($J$),
  and  slope ($L$) of the symmetry energy at  $n_0$. The last line
  shows, for comparison, these quantities for the
  APR EoS~{\protect\citep{APR}.}
}
\label{tab:models}
\end{table}
\begin{table*}
\setlength{\tabcolsep}{4.2pt}
\begin{tabular}{lc c cc cc cc cc cc c cc}
\hline
Model &  mesons & flavor & $n_{\rm max}$ & $M_{\rm max}^{ Y}$ & $Y_1$ & $n_{Y_1}$ & $M_{Y_1}$     &
$Y_2$ & $n_{Y_2}$ &  $M_{Y_2}$  & $Y_3$  & $n_{ Y_3}$ &  $M_{Y_3}$  & $n_{\rm DU}$ & $M_{\rm DU}$ \\
      &         & sym.   & [fm$^{-3}$]   & [$M_{\odot}$]      &      & [fm$^{-3}$]& [$M_{\odot}$] &
      &[fm$^{-3}$]&[$M_{\odot}$]&        & [fm$^{-3}$]& [$M_{\odot}$]& [fm$^{-3}$] &[$M_{\odot}$] \\
\hline
DDME2 & $\sigma$, $\omega$, $\phi$, $\rho$                         & SU(6)  & 0.93 & 2.12 &
$\Lambda$ & 0.34 & 1.39 & $\Xi^-$    & 0.37 & 1.54 &$\Sigma^-$& 0.39 & 1.60 & $-$    & $-$    \\
SWL   & $\sigma$, $\omega$, $\rho$                                 & SU(3)  & 0.97 & 2.00 &
$\Lambda$ & 0.41 & 1.51 & $\Xi^-$    & 0.45 & 1.65 & $\Xi^0$  & 0.90 & 2.00 & 0.90 & 2.00 \\
GM1A  & $\sigma$, $\omega$, $\phi$, $\rho$                         & SU(6)  & 0.92 & 1.99 &
$\Lambda$ & 0.35 & 1.49 & $\Xi^-$    & 0.41 & 1.67 & $-$        & $-$    & $-$    & 0.28 & 1.10 \\
PKA1  & $\sigma$, $\sigma^\ast$, $\omega$, $\phi$, $\rho$, $\pi$   & SU(3)  & 1.09 & 2.01 &
$\Lambda$ & 0.32 & 1.43 & $\Sigma^-$ & 0.39 & 1.57 &  $\Xi^-$ & 0.43 & 1.63 & 0.25 & 0.98 \\
PKO1  & $\sigma$, $\sigma^\ast$, $\omega$, $\phi$, $\rho$, $\pi$   & SU(3)  & 0.96 & 2.02 &
$\Lambda$ & 0.33 & 1.52 &  $\Xi^-$   & 0.58 & 1.87 & $-$        & $-$    & $-$    & 0.25 & 1.01 \\
PKO2  & $\sigma$, $\sigma^\ast$, $\omega$, $\phi$, $\rho$          & SU(3)  & 0.95 & 2.03 &
$\Lambda$ & 0.34 & 1.55 &  $\Xi^-$   & 0.50 & 1.86 & $-$        & $-$    & $-$    & 0.30 & 1.28 \\
PKO3  & $\sigma$, $\sigma^\ast$, $\omega$, $\phi$, $\rho$, $\pi$   & SU(3)  & 0.97 & 2.02 &
$\Lambda$ & 0.33 & 1.55 & $\Sigma^-$ & 0.45 & 1.81 & $\Xi^-$  & 0.48 & 1.84 & 0.28 & 1.22 \\
\hline
\end{tabular}
\caption{Astrophysical characteristics of the relativistic DF EoS
  models (with hyperons) used in this work: $n_{\rm max}$ shows the
  central densities of the maximum-mass ($M_{\rm max}^Y$) hyperonic
  star of each stellar sequence, $n_{Y_i}$ shows the threshold
  densities at which hyperons of type $Y_i$ are produced, and
  $M_{Y_i}$ lists the mass of the hyperonic star for that
  central density.  The last two
  entries show the baryon number density ($n_{DU}$) beyond which the
  nucleonic dUrca process is allowed in purely nucleonic NS matter
  and the mass ($M_{DU}$) of the associated compact star.}
\label{tab:rmfY}
\end{table*}

The EoS of hypernuclear matter used in the present study are displayed
in the upper panel of Fig.~\ref{fig:EOS}. For comparison, we also show
in this figure the purely nucleonic non-relativistic microscopic EoS
of \cite{APR} (APR) which is frequently used in cooling simulations of
nucleonic NSs. It is seen that the EoS predicted by APR and the DDME2
  and SWL models are close to each other for $n \lesssim n_0$,
  where the low-$L$ mean-field models reproduce the results of pure neutron matter (PNM)
  obtained by ab initio models ~\citep{Fortin_PRC2016}.
  In the density
range $n_0 \lesssim n\lesssim 4n_0$ the APR EoS is softer than the
relativistic DFT based EoS; at higher densities the inverse is
true. The latter feature is mainly due to the fact that the
relativistic DFT based EoS feature increasingly abundant hyperonic
components at higher densities, which softens the EoS. A strict
comparison between the models can be carried out in the case of PNM,
in which case the contribution of the iso-vector sector to the EoS of
different models becomes clear. At densities $n \gtrsim 2n_0$
the value of the energy per neutron
$(E/A)^{\rm (PNM)}$ predicted by the DFT models
is larger than that predicted by APR. Taking as an example the DDME2
model we find that this quantity is larger by 35$\% $ at $2 n_0$,
87$\% $ at $4 n_0$ and 61$\%$ at $6 n_0$.  The EoS stiffness has
important consequences for the theoretically established pressure and
energy density values at the centers of compact stars.  For instance,
as shown in Fig.~\ref{fig:EOS}, for a $1.4\, M_{\odot}$ star
the central energy density of APR is by a factor of 1.36 to 1.73 larger
than the energy densities obtained for our EoS.

The mass-radius and mass--central-density dependencies are shown in
the middle and bottom panels of Fig.~\ref{fig:EOS}.  As can be seen,
all models considered in our study satisfy the $M\simeq 2\, M_{\odot}$
NS mass constraint.  The GM1A and PK models have $P(n_B)$ dependences
and $L$ values close to each other so that the neutron stars with high
enough ($n_c\ge 3n_0$) central densities have similar masses and
radii.  The differences just above the saturation density are
reflected in variations of the radii of these models. The relativistic
Hartree models DDME2 and SWL predict radii of the order of 13 km,
while the HF based models tend to predict larger radii, on the order
of 14 km.  The recent detection of gravitational waves from GW170817
places an upper limit on the NS tidal deformability, and provides
evidence that the radii of the merging NSs was $R \le 13.7$
km~\citep{Abbott2017,Paschalidis2017,Fattoyev2018,Annala2018,Most_PRL_2018,Christian2019,Tews2019}
with some uncertainty in the masses. The total mass of the binary is
in the range $2.73\le M/M_{\odot}\le 2.78$, which is split among the
members of the binary in the range $1.36 \le M/M_{\odot}\le 1.6$ for
one star and $1.17 \le M/M_{\odot}\le 1.36$ for the other.  Note that
the limit on the radius is extracted from the measured tidal
deformability.  The value of the radius is clearly related to the
value of $L$, where smaller radii correspond to smaller $L$ values
(see Table~\ref{tab:rmfY}).  Such correlation was observed on
  a much larger set of phenomenological purely nucleonic EoS by
  \cite{Fortin_PRC2016}, who find that the correlation is stronger in
  less massive stars, which cover densities not much larger than
  $n_0$.  Such correlation is confirmed by the fact that PK and GM1A
  models with their large $L$-values predict large radii. While these
  models overshoot the limit placed by the GW170817 analysis by about
  $10\%$,  this discrepancy is not significant in the
  context of cooling. Indeed, the radius enters explicitly only the photon
  luminosity from the surface of the star which is unimportant for our
  discussion; the variations in the internal composition and
  distribution of particles over the internal radius are small,
  leading to $10\%$ change in the radius value.  The bottom panel of
Fig.~\ref{fig:EOS} shows that those EoS which are soft up to densities
$\lesssim 4 n_0$ produce intermediate mass NS with high-density cores.

A more quantitative overview of the properties of hypernuclear neutron
stars is given in Table \ref{tab:rmfY} for the EOS of this work.  The
central density of the maximum mass star (fourth column), the value of
the maximum mass (fifth column) and the density at which the first
hyperon species appears (columns 6 and 7) are remarkably close to each
other (exceptions are $n_{\rm max}$ for PKA1 and $n_{Y1}$ for the SWL
model). The differences in the models are larger for the remaining
entries of the table, as these are controlled by the high-density
behavior of the EoS. An important difference concerns the onset of the
nucleonic dUrca process (column 15 of Table~\ref{tab:rmfY}). DDME2
does not allow for this process in stable stars.  The SWL models only
allow for it at very high densities ($n\ge 0.9$ fm$^{-3}$).  Finally
the other models predict considerably lower dUrca thresholds
($\sim 0.3$ fm$^{-3}$). We also note the (negative) correlation
between the value of $L$ and the threshold density of the Urca process
(column 15 of Table~\ref{tab:rmfY}). For purely nucleonic models, such
correlation was investigated by \cite{Fortin_PRC2016} on a larger set
of density functional models which produce $2M_{\odot}$ NS. These
authors find that: (i) for $L \geq 60$ MeV, there is a clear
anti-correlation between $L$ and $n_{DU}$, and (ii) for $L < 60$ MeV the
situation is ambigous as some models allow the dUrca to operate in stars
as small as $1.5M_{\odot}$ while others forbid it in all stable
configurations.  Clearly, other factors play in this situation a
decisive role. 

\begin{figure}
\begin{center}
\includegraphics[angle=0, width=0.99\columnwidth]{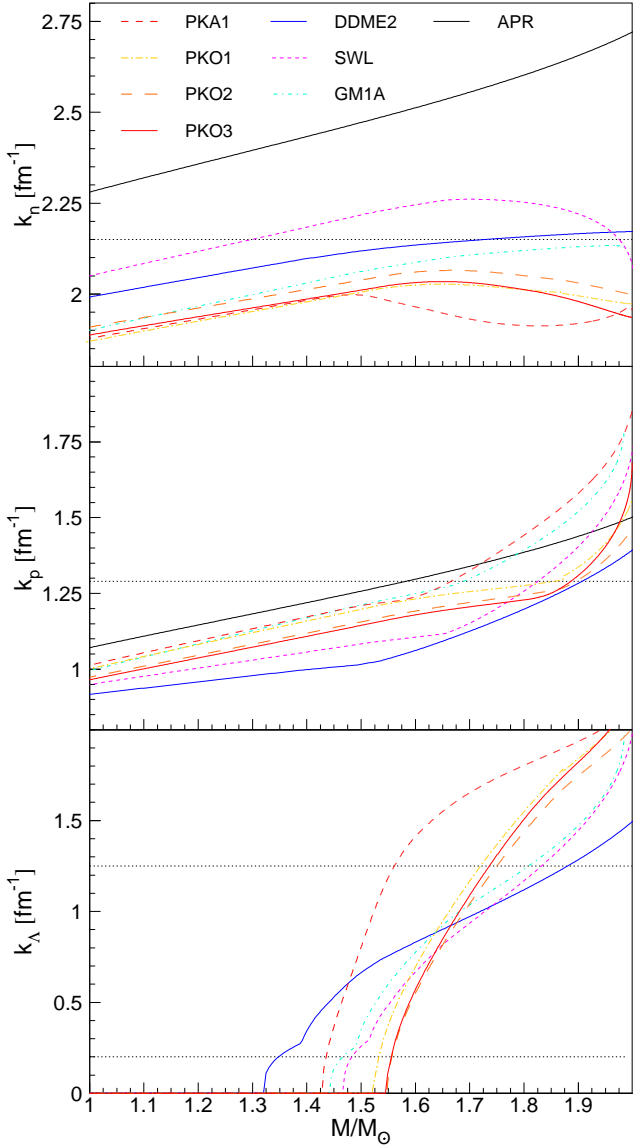}
\end{center}
\caption{Fermi momenta of neutrons (top), protons (middle) and
  $\Lambda$ (bottom) in the centers of NS of different masses, for the
  collection of EoS of this work. The horizontal dotted lines mark the
  upper limits (top and middle panel) or the boundaries (bottom panel)
  of the momentum range for which neutron, proton and $\Lambda$
  pairing occurs. Neutron $P$-wave pairing results are from
  {\protect\cite{Ding_PRC_2016}}. For proton $S$-wave pairing we used
  results from {\protect\cite{Chen_NPA1993}}. The $\Lambda$ $S$-wave
  pairing is obtained from the BCS equation (see text for details).}
\label{fig:kF_NSM}
\end{figure}

Furthermore, it is important to discuss the particle content of the
models under consideration, as the thermal evolution significantly
depends on the composition of matter.  Rather than showing the
abundances of particles [see, for instance, Paper~I and
\cite{Li2018EPJA,Li2018PLB}], we show in Fig.~\ref{fig:kF_NSM} the
Fermi momenta of the most abundant particle species, neutrons (top),
protons (middle) and $\Lambda$ (bottom).  We also show the range of
momenta over which neutron $P$-wave pairing \citep{Ding_PRC_2016}, and
proton $S$-wave pairing \citep{Chen_NPA1993} and $\Lambda$ $S$-wave
pairing \citep{Raduta2018MNRAS} occur.  Thus, the figure shows in a
transparent manner the existence of various condensates in the centers
of NSs.  The suppression of pairing in any one of these channels leads
to significantly faster stellar cooling.

\begin{figure}
\begin{center}
\includegraphics[angle=0, width=0.99\columnwidth]{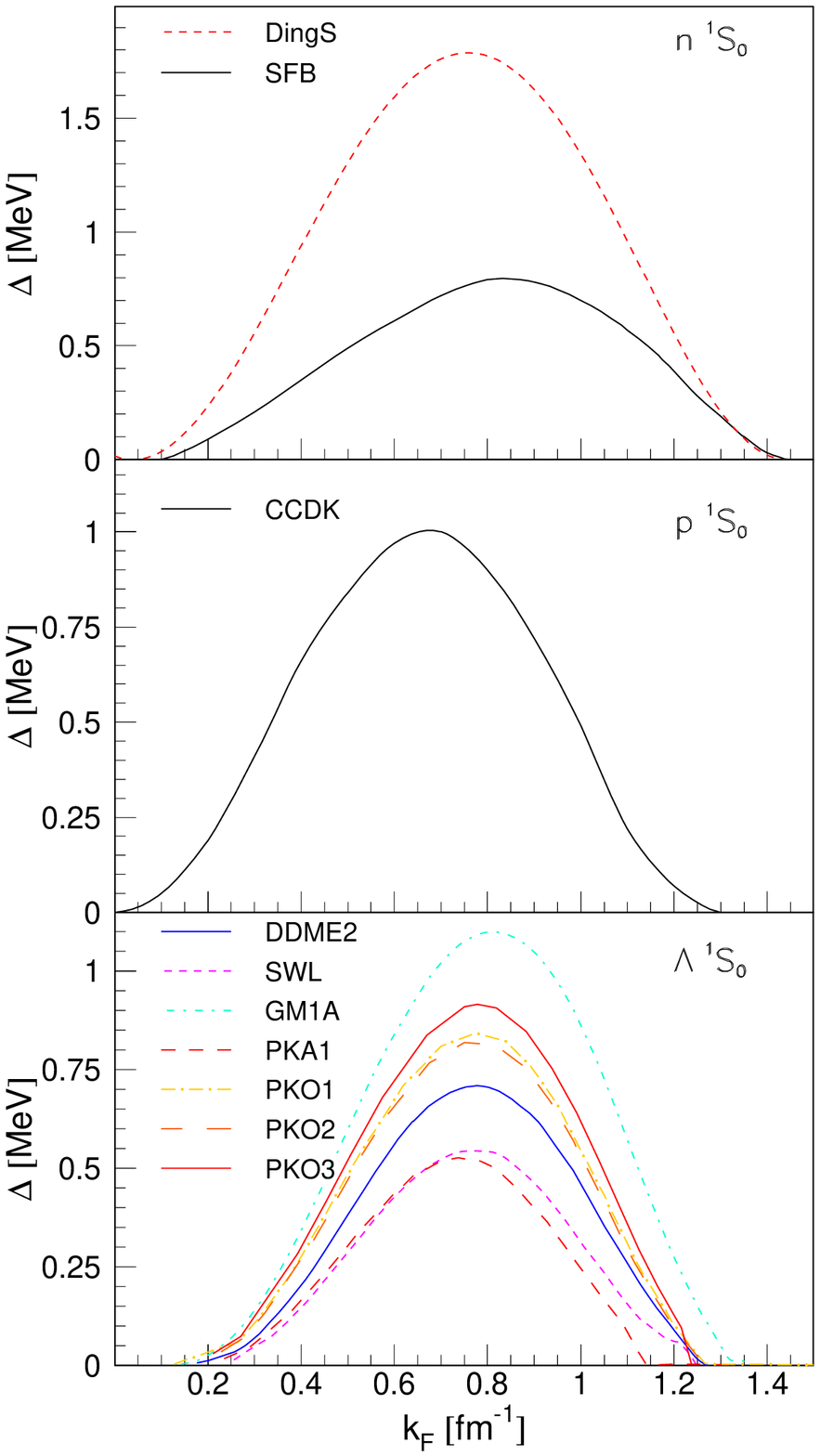}
\end{center}
\caption{{$S$}-wave pairing gaps for neutrons (top), protons (middle)
  and $\Lambda$'s (bottom) as a function of Fermi momentum.  The SFB
  neutrons gaps \citep{SFB_2003} include long-range correlations only,
  whereas long- and short-range correlations are included in the DingS
  gap \citep{Ding_PRC_2016}.  For protons, we show the $^1S_0$
  gap as computed by \protect\cite{Chen_NPA1993}. The $\Lambda$
  $^1S_0$ gaps are obtained as in Paper I with the bare
  $\Lambda$-$\Lambda$ interaction taken to be the configuration space
  parametrization of the ESC00 potential~\citep{2001NuPhA.691..322R}
  given by {\protect\cite{Filikhin_NPA2002}}. The composition of
  matter itself and the $\Lambda$ single-particle spectra are computed
  for the EoS listed in the bottom panel.  }
\label{fig:swave_gaps}
\end{figure}

It has been known for a long time that the cooling simulations
  are sensitive to the choice of the gaps of fermionic constituents of
  NS matter, for an overview see ~\citep{Sedrakian2018}. We now
  comment briefly on the gap choices made in this work. For the
  neutron $^3P_2$-$^3F_2$-wave gap we have implemented the results of
  \cite{Ding_PRC_2016}, specifically those that have been obtained
  using the Argonne v18 \citep{Wiringa_PRC_1995} interaction.  This
  calculation includes both the short- and long-range correlations in
  neutron matter as well as three-body contributions. Because the
  $^3P_2$-$^3F_2$ pairing occurs at high densities the bare
  interaction is not constrained by the phase-shift analysis and the
  results corresponding to different interactions may differ in the
  high-density segment relevant for neutron stars. In addition, the
  three-body force produces a non-negligible effect. The many-body
  scheme used by \cite{Ding_PRC_2016} is one of the most reliable to
  date. We will also use alternatively the result obtained within the
  BCS scheme with bare two-body interaction in combination with a
  phase-shift equivalent interaction~\citep{Baldo_PRC58}.  For the
  proton $S$-wave gaps we use the results of \cite{Chen_NPA1993} which
  are based on the matrix elements extracted from Reid-soft-core
  potential~\citep{Chao_NPA_1972}. Our choice, in this case, is
  motivated by the fact that the corresponding gap, in its magnitude
  and density range, is the largest and broadest in the literature,
  i.e., it maximizes the effect of proton pairing on the physics
  studied.  In particular, this choice very efficiently slows down the
  cooling induced by $(\Lambda,p)$ dUrca process [see Paper I for
  extensive discussion].  Finally, $\Lambda$ $S$-wave pairing gaps
  where obtained by solving the BCS equation for hyperonic pairing, as in
  Paper I.  For more details on $\Lambda$-pairing, see the comments on
  Fig.~\ref{fig:swave_gaps}.  

As shown in the top panel of Fig.~\ref{fig:kF_NSM}, the neutron Fermi
momenta in the centers of NSs computed for the low-$L$ EoS
of our collection, i.e., DDME2 and SWL, are larger than the Fermi
momenta obtained for the large-$L$ EoS. This follows from the fact
that the former EoS produce stars that are more compact (for the same
mass). For example, we find that for a $M/M_{\odot}=1.0$ mass star the
central neutron densities obtained for the low-$L$ EoS are
$n_{c,n}^{\rm DDME2}=0.26$ fm$^{-3}$ and $n_{c,n}^{\rm SWL}=0.29$
fm$^{-3}$, whereas for {the} large-$L$ EoS
$n_{c,n}^{\rm PK} \simeq 0.22$ fm$^{-3}$. The same values for { a}
$M/M_{\odot}=1.8$ mass star are $n_{c,n}^{\rm DDME2}=0.33$ fm$^{-3}$,
$n_{c,n}^{\rm SWL}=0.38$ fm$^{-3}$ and
$0.23 \leq n_{c,n}^{\rm PK} \leq 0.29$ fm$^{-3}$.  Note that the
neutron abundance does not increase necessarily with the mass of the
star (which is the case for DDME2 and GM1A), as the particle
abundances are regulated in a convoluted manner by the conditions of
$\beta$-equilibrium and charge neutrality.  The range of momenta below
which neutron $^3P_2$-$^3F_2$ superfluidity exists according to
\cite{Ding_PRC_2016} when neutron-neutron interaction is described by
the Argonne v18 \citep{Wiringa_PRC_1995}
is marked by a dotted line (the lower limit is at
1.15 fm$^{-1}$ and is outside the figure's scale).  It is seen that
neutron $^3P_2$-$^3F_2$-wave superfluidity always extends to the center of a star
of given mass for all large-$L$ EoS, whereas massive stars based on
DDME2 ($M/M_{\odot} \geq 1.7$) and SWL
($1.3 \le M/M_{\odot} \le 1.95$) may feature normal neutron fluids in
their centers. We also note that the large difference between the DFT
based EOSs and the APR EOS is inherent in these models and not
attributable to the fact that the APR EoS does not include
high-density hyperonic components.

The middle panel of Fig.~\ref{fig:kF_NSM} shows the proton Fermi
momentum at the star's center for our collection of EoS, which
increases with the mass of the NS.  The variations in the
proton fraction reflects the uncertainty in the behavior of the
symmetry energy away from the saturation density.  Not surprizingly, a
correlation is observed between the $L$ value and the proton fraction
in the cores of NSs, such that the EOS with the lowest $L$
value leads to the lowest proton abundance in the core. It is also
remarkable that all models show a change in the slope, which reflects
a rapid increase of proton fraction as the mass increases. The upper
limit {on} the momenta where proton $S$-wave superfluidity exists
according to \cite{Chen_NPA1993} is shown by the horizontal line.
Clearly, the low-mass stars feature proton $^1S_0$ superfluidity 
which extends all the way to the stellar center, whereas in more
massive stars, with $M/M_{\odot} \ge 1.6$, the  $S$-wave proton pairing
vanishes at some density which is below the central density. 
This is a crucial feature that accelerates the cooling of massive stars.
We will return to this problem below.

The bottom panel in Fig.~\ref{fig:kF_NSM} shows the Fermi momentum of
$\Lambda$ hyperons. The DDME2 and SWL EoS predict low $\Lambda$
fractions in massive stars, but there is no clear distinction between
the EoS at intermediate stellar masses. The momentum range for which
$\Lambda$ $S$-wave superfluidity exists is bracketed by horizontal
lines. Thus, we observe that the intermediate mass stars contain a
$\Lambda$ $S$-wave superfluid over the whole volume, whereas the most
massive stars have cores containing either normal $\Lambda$'s or
$\Lambda$'s paired through a higher partial wave, as discussed below.

Figure \ref{fig:swave_gaps} shows $S$-wave pairing gaps for neutrons,
protons, and $\Lambda$'s, which have been used in our cooling
simulations.  The two models used for neutron $S$-wave pairing reflect
the uncertainties in the $S$-wave gap related to treatment of
correlations among neutrons.  The maximum value of the
  $S$-wave gap when both short- and long-range correlations are taken
  into account \citep{Ding_PRC_2016} is more than twice as large as
  the maximum value obtained when only long-range correlations
  (polarization effects) are included \citep{SFB_2003}.  Because the
  neutron-neutron interaction at low energies is constrained by the
  scattering data the main uncertainty in the value of the gap arises
  from the many-body treatment and not from the bare interaction.
  See~\cite{Sedrakian2018} for a recent discussion.   The main effect
of neutron $S$-wave pairing in the crust is to regulate the
thermalization of the crust. In the absence of the nucleonic dUrca
process, a larger gap in the crust enhances the cooling rate.  The
opposite effect occurs if the nucleonic dUrca operates.  Neutron
pairing in the crust impacts the cooling mainly over a time span of
$ t\lesssim 10^2$ yr, during which the star is not yet isothermal and
the surface temperature is controlled by the crust.  Therefore, the
variations in the gap value are not significant for the interpretation
of the data. The proton $S$-wave pairing gap, which is non-zero in the
star's core, plays an important role for regulating dUrca processes
involving protons. In particular, this pairing gap regulates the
$\Lambda p$ channel of the hyperonic dUrca process, as discussed in
Paper I and by \cite{Grigorian_NPA_2018} and \cite{Tolos2018ApJ}.
Here we adopt the proton gap of \cite{Chen_NPA1993} which covers the
largest density range and has a relatively large maximum of the order
of 1 MeV.  Thus, the suppression effects through the proton $S$-wave
gap will be maximized in our cooling simulations.

We have (re)computed the $\Lambda$ $^1S_0$ pairing gaps within the
standard BCS theory, as described in Paper I.  As in Paper I, for the
bare $\Lambda$-$\Lambda$ interaction we used the configuration space
parametrization of the ESC00 \citep{2001NuPhA.691..322R} provided by
\cite{Filikhin_NPA2002}.  In computing the gaps we use the background
matter composition and single particle spectrum of $\Lambda$'s in the
normal state provided by the EoS under consideration.  The
EoS-dependence of the gap enters via the Dirac effective mass.  The
reduction of the hyperon masses by medium effects diminishes the
density of states and, thus, the gap at the Fermi surface.  ESC00
\citep{2001NuPhA.691..322R} gives for $\Lambda$-$\Lambda$ one of the
most attractive interactions in the literature. The suppression of
dUrca processes involving $\Lambda$'s is thus maximized. Note that the
$\Lambda$ $S$-wave superfluidity occurs for Fermi momenta
$0.2\le {k_{F_\Lambda}} \le 1.3$ fm$^{-1}$, which is about the same
range as for $S$-wave neutron and proton condensates (see
Fig.~\ref{fig:swave_gaps}).  It is also seen that the gaps are in the
range $0.5\le \Delta_{\Lambda} \le 1.2$ MeV, which is comparable to
the proton pairing gap. We thus anticipate a competition of these two
pairing channels when considering the suppression of the $\Lambda p$
channel of the hyperonic Urca process.

\begin{figure}
\begin{center}
\includegraphics[angle=0, width=0.99\columnwidth]{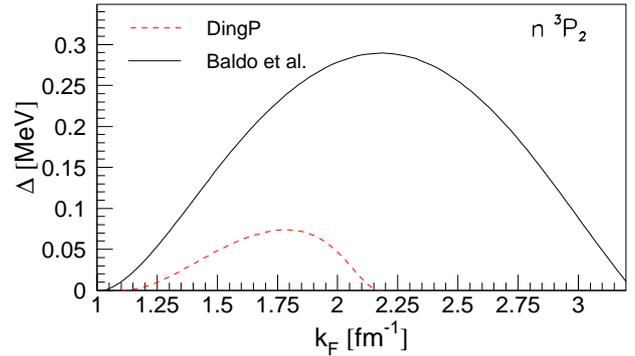}
\end{center}
\caption{Neutron $^3P_2$-$^3F_2$ pairing gaps as a function of
    Fermi momentum.  The results of \protect\cite{Ding_PRC_2016} which 
includes short- and long-range correlations is shown by the dashed
line, the result of  \protect\cite{Baldo_PRC58} which is based on BCS
theory is shown by the solid line. 
}
\label{fig:pwave_gaps}
\end{figure}

The $P$-wave pairing of neutrons is harder to compute as the
interactions are not constrained by the phase-shift analysis in the
high-energy domain needed for this channel. To illustrate the range of
theoretical possibilities we present two models of the $^3P_2$-$^3F_2$
pairing gap, which are shown in Fig.~\ref{fig:pwave_gaps}.
The first choice corresponds to computations of \cite{Ding_PRC_2016},
already discussed in connection with Fig. \ref{fig:kF_NSM}.
They use the Argonne v18 \citep{Wiringa_PRC_1995} interaction
and account for short- and long-range correlations.
The gap covers the Fermi-momentum range $1.15 \le {k_{F}} \le 2.1$ fm$^{-1}$.
Its maximum, $\Delta \simeq 0.074$ MeV, occurs for $k_F=1.8$ fm$^{-1}$.
This calculation is taken as a lower bound on the gap.
As an upper limit on the $^3P_2$-$^3F_2$-wave pairing gap we adopt the result of
\cite{Baldo_PRC58}, which corresponds to the CD Bonn \citep{CD-Bonn}
interaction and the Bruckner-Hartree-Fock spectrum.
The gap covers a significantly broader range of momenta.
The maximum of this gap ($\sim$ 0.3 MeV at $k_{F}\sim 2.15$
fm$^{-1}$) is a factor of four larger than the gap of \cite{Ding_PRC_2016}.
In passing we note that a number of studies of
the cooling of NSs assumed a {\it vanishing} $^3P_2$-$^3F_2$ pairing
gap [see \cite{Beznogov_PRC_2018,Grigorian_NPA_2018,Tolos2018ApJ}].
This assumption improves the agreement of the cooling curves with the
data on surface temperatures of some compact objects, but it may lack
any theoretical underpinning.  In one of the scenarios considered in
this paper, we shall make this assumption as well.

\subsection{BCS $P$-wave pairing for protons and $\Lambda$'s}
\label{sec:pairing}

We have previously addressed the Bardeen-Cooper-Schrieffer (BCS)
pairing among hyperons in the low-density limit, where the most
attractive partial wave is the $^1S_0$ channel. In this case hyperons
are paired in spin-singlet Cooper pairs. Note that within the
hypernuclear mixture of particle, pairing among the different species
is strongly disfavored by the particle mass differences (except for
neutrons and protons, in which case the pairing is suppressed by their
large difference in number density).

The $^1S_0$ gaps for $\Lambda$ hyperons were computed using the
non-relativistic BCS equation for a given hyperon-hyperon potential,
whereas the single-particle energies and particle composition were
computed for the relativistic DF method, as described in Paper I. [See
also the related earlier work on hyperonic pairing by
\cite{Balberg_PRC1998}, \cite{Wang_PRC2010} and
\cite{Takatsuka_XiPairing}.] Computing the single-particle energies
and pairing gaps for different theories is not consistent formally,
but in practice such an approach has been proven to work well for
finite nuclei~\citep{Vretenar2005,Kucharek1991ZPhyA}. It is also in
line with the so-called decoupling approximation which has been
applied to the many-body treatment of nuclear pairing~ [for a
review, see~\citep{Sedrakian2018}].  We recall that for the 
$\Lambda\Lambda$ pairing interaction we used the
configuration-space parameterization of the ESC00
potentials~\citep{2001NuPhA.691..322R} given by
\cite{Filikhin_NPA2002}.  
The potential was chosen in such a manner as
to maximize the pairing gap, i.e., to provide an upper limit on the
values of the gap function.  Note also that the $\Sigma \Sigma$ pairing was 
disregarded, though this possibility has been previously considered
\citep{Vidana_2004PhRvC} and gaps as large
 as several MeV have been obtained.

On the basis of the (approximate) isospin invariance of nuclear
forces, it is natural to anticipate that $^3P_2$-$^3F_2$ pairing
should also occur in the proton component, for the range of densities
where this channel dominates the attraction among protons.  To get an
estimate of pairing gap of this channel, we account for the difference
between the density of states of neutrons and protons, which amounts
to replacing the neutron Landau mass $m_n^*$ by the proton Landau mass
$m_p^*$. This implies a rescaling of the dimensionless pairing
interaction by the factor $\alpha_p^{-1} = m^*_p/m^*_n$.  The result
is the weak-coupling estimate for the $P$-wave pairing gap for protons
as
\begin{eqnarray}
\label{eq:gap_ratio_p}
\Delta_{p} = \epsilon_{{F_p}}
 \left(\frac{\Delta_n}{\epsilon_{{F_n}}}\right)^{\alpha_p},
\end{eqnarray}
where $\epsilon_{{F_i}}$ is the Fermi energy of baryon $i$. Thus, from
the $P$-wave pairing gap for neutrons and the Fermi energies of
particles, we obtain an estimate of the $P$-wave pairing gap for
protons.  Equation \eqref{eq:gap_ratio_p} follows directly from the
BCS formula
$
\Delta_{i} = \epsilon_{Fi}  \exp[-1/(\nu_i V_i)],
$
where $\nu_i$ and $V_i$ are { the} density of states and the pairing 
matrix element for a baryon of type $i$.  

We further speculate about $P$-wave pairing among $\Lambda$'s.  An
estimate of this pairing gap follows from the strategy adopted for
protons, except that now we additionally rescale the pairing
interaction among neutrons by a factor of $2/3$ according to the
SU$(6)$ flavor quark model.  This means that the dimensionless pairing
interaction is rescaled by the factor
$\alpha_{\Lambda}^{-1} = 2m_\Lambda^*/3m_n^*$ and, therefore, 
\begin{eqnarray}
\label{eq:gap_ratio_Lambda}
\Delta_{\Lambda} = \epsilon_{{F_\Lambda}}
 \left(\frac{\Delta_n}{\epsilon_{F_n}}\right)^{\alpha_{\Lambda}}.
\end{eqnarray}
\begin{figure}
\begin{center}
\includegraphics[angle=0, width=0.99\columnwidth]{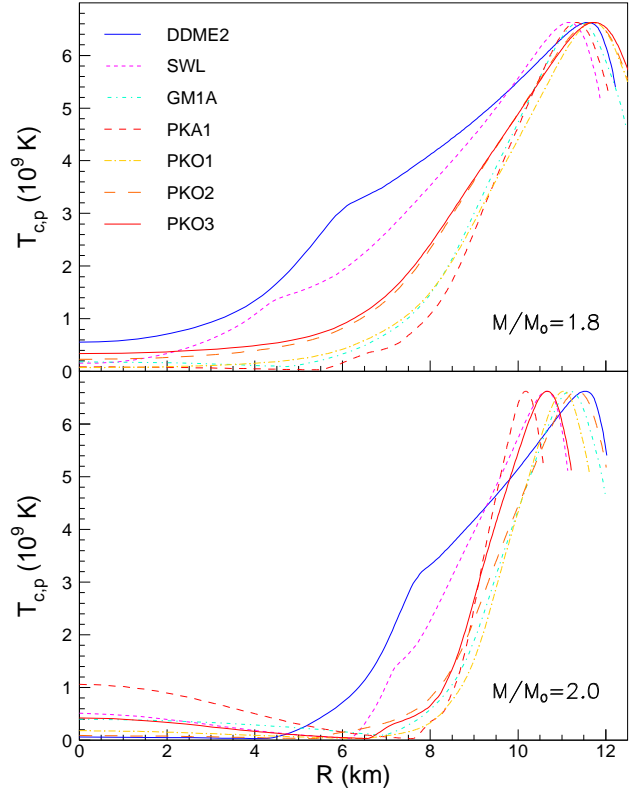}
\end{center}
\caption{Critical temperature of proton superfluidity for 
    NSs with masses of $1.8M_{\odot}$ (top) and
  $2M_{\odot}$ (bottom){, for the EoS of this work}.  $^1S_0$ pairing
  is { modeled} as in the paper of CCDK \citep{Chen_NPA1993}.  The
  proton $P$-wave $T_c$ is obtained by 
re-scaling the neutron $P$-wave
  gap, given in {\protect\cite{Baldo_PRC58}}, via
  Eq.~(\ref{eq:gap_ratio_p}).}
\label{fig:Tcp}
\end{figure}
\begin{figure}
\begin{center}
\includegraphics[angle=0, width=0.99\columnwidth]{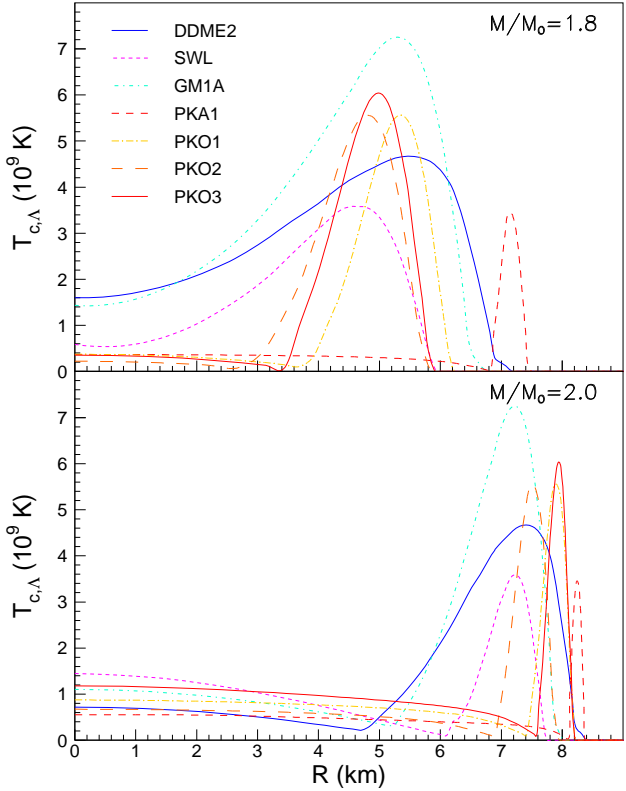}
\end{center}
\caption{Critical temperature { of} $\Lambda$ superfluidity
  for stars with masses of $1.8M_{\odot}$ (top) and $2M_{\odot}$
  (bottom), for the EoS of this work.  The $^1S_0$
  pairing is treated in the BCS theory, as in Paper I. 
    The critical temperature of the $P$-wave superfluid is obtained by
    re-scaling the neutron $P$-wave gap \citep{Baldo_PRC58} via
    {Eq.} (\ref{eq:gap_ratio_Lambda}). }
\label{fig:TcL}
\end{figure}
Figures~\ref{fig:Tcp} and \ref{fig:TcL} show the critical temperature
profiles for protons and $\Lambda$'s inside of NSs with masses of
$1.8 M_{\odot}$ (top) and $2 M_{\odot}$ (bottom) built using different
EoS.  For the proton $^1S_0$-pairing we adopt the results of
\cite{Chen_NPA1993}.  The $\Lambda$ $^1S_0$-pairing is determined, for
each EoS, by solving the BCS equation, as explained before.  The
$^3P_2$-$^3F_2$-pairing gap of protons and $\Lambda$'s are obtained by
the re-scaling procedure described just above, using
Eqs.~\eqref{eq:gap_ratio_p} and \eqref{eq:gap_ratio_Lambda}.  In both
cases, the results of \cite{Baldo_PRC58} are used for the neutron
$^3P_2$-$^3F_2$-pairing.  High density pairing gaps based on
  the $^3P_2$-$^3F_2$ gap by \cite{Ding_PRC_2016} (not illustrated)
  feature similar or lower critical temperatures in the innermost
  core, depending on EoS and NS mass.  The double-bell shaped
structures arise, in the case of $2 M_{\odot}$ NS and, for some EoS,
also $1.8 M_{\odot}$ NS, because of the two ($S$- and $P$-wave)
pairing channels, each channel giving rise to a maximum. The maximum
appearing for large internal radius corresponds to the $S$-wave
pairing, whereas the one appearing for small internal radius - to the
$P$-wave pairing.  Very often and especially in the most massive
stars, there is a shell, between those featuring $S$- and $P$-wave
condensates, where pairing almost vanishes.  It is straightforward to
anticipate that its role in cooling will be significant.

Because of their low density, one could expect that the protons manifest
$^1S_0$ superfluidity in the entire stellar core.
Fig. \ref{fig:Tcp} shows that this is not true.
For massive stars and most EoS protons are superfluid only in the outer core.
When they are additionally allowed to pair in the $^3P_2$-$^3F_2$ channel,
the situation does not improve much as, very often, in the inner most regions
the critical temperature is so low that it falls below the stellar temperature.
In this event proton superfluidity will not be experienced and
it will play no role in the suppression of neutrino radiation from the stellar core.
Thus, while $P$-wave pairing of protons could in principle provide an
efficient mechanism for the suppression of neutrino radiation,
its actual occurrence depends on the adopted EoS and on
the density range covered by the mass model. Note that the critical
temperature for the $S$-wave pairing has a maximum value of
$T_{cp}\simeq 7\times 10^9$ K, whereas for $P$-wave pairing this is of
the order of $10^9$~K or less.  We conclude that for most of the EoS
of this study there exist domains, either in the inner core or at the
boundary between inner and outer core, where $T_{c,p}$ is too low to
efficiently suppress cooling via dUrca process involving protons.

For both NS masses considered in Fig.~\ref{fig:TcL} $\Lambda$ hyperons are in
the $S$-wave superfluid state in the outer core.
For PKA1, which predicts the highest $\Lambda$ densities, the superfluid shell is very narrow.
$T_{c,\Lambda}$ ranges in the interval $3 -7\times 10^9$~K and strongly depends on the EoS,
the star's mass and the distance from the center.
For DDME2 and GM1A $S$-superfluidity is encountered also in the inner core
of the $1.8M_{\odot}$ star.
$\Lambda$ $P$-wave pairing leads to critical temperatures higher than those
obtained for protons.
The radial changes of $T_{c,\Lambda}$ can be attributed to the different fractions
and effective masses of the $\Lambda$ particle, as predicted by different EoS.
For a large enough $\Lambda$ population we find a $P$-wave $\Lambda$ superfluid in the
inner core for a number of models.  The values of $T_{c,\Lambda}$ are
typically between $2 \times 10^8$ and $1.5 \times 10^9$ K.  

In conclusion, it is clearly seen that in the absence of proton and
$\Lambda$ $P$-wave pairing, NSs cool rapidly (as already demonstrated in Paper I)
via the dUrca process $\Lambda \to p + l + \bar\nu_l$, where $l$ stands for a lepton.
This will be different if high-density $P$-wave pairing of protons and
$\Lambda$'s occurs, which suppresses the dUrca process (see
Sec.~\ref{sec:therm_ev} below for explicit examples).

\section{Thermal evolution of hypernuclear stars}
\label{sec:therm_ev}

We have performed a large number of cooling simulations of stellar
models based on our EoS collection of hypernuclear stars described
above.  We assume spherically symmetric configurations of isolated
self-gravitating NSs which are non-rotating and
non-magnetized.  Heating sources were not considered.  Our
collection of EoS was supplemented with the crust EoS of
\cite{NEGELE1973298} and \cite{1989A&A...222..353H}, which is smoothly
joined together with the core EoS. The outer envelope of the star
is assumed to consist of Fe. Note that young stars are believed to
have He or C atmospheres.  Light-element envelopes lead to surface
temperatures which are higher than those produced by heavy-element
envelopes. This means that the use of a Fe envelope throughout the
entire evolution artificially diminishes the effective temperature of
young stars and partially compromises the agreement with the X-ray
data on their thermal emission.

Our models were evolved in time from an arbitrary radial temperature profile
under the assumption that the structure of the models does not change in time.
This initial distribution is such that the temperature smoothly
  decreases from $10^{10}$ K in the center to $8 \cdot 10^9$ K at the boundary between
  the inner and outer crust and, finally, to $5 \cdot 10^9$ K at the surface.

As in Paper I, we employ an extension of
the public domain NSCool
code\footnote{www.astroscu.unam.mx/neutrones/NSCool} to include the
hyperonic component with its radiation processes.  These include the
hyperonic dUrca processes~\citep{1992ApJ...390L..77P}
\begin{eqnarray} 
\label{eq:UrcaLambda}
\Lambda &\to& p + l  + \bar\nu_l,\\
\label{eq:UrcaSigmaminus}
\Sigma^- &\to& \left(\begin{array}{c} n  \\
                       \Lambda \\
\Sigma^0
\end{array} \right) + l + \bar\nu_l,\\
\label{eq:UrcaXiminus}
\Xi^- &\to& \left(\begin{array}{c} \Lambda  \\ 
                                                         \Xi^0
                        \\
\label{eq:UrcaSigmazero}
\Sigma^0   \end{array} \right) + l + \bar\nu_l,\\
\label{eq:Xizero}
\Xi^0 &\to& \Sigma^+ + l  + \bar\nu_l,
\end{eqnarray}
where $l$ stands for a lepton, either electron or muon, and
$\bar\nu_l$ is the associated anti-neutrino.  In addition, we include
the neutrino emission via the Cooper pair-breaking and formation (PBF)
mechanism:
\begin{eqnarray}
\label{eq:Y_PBF}
\{YY\} \to Y+Y + \nu + \bar \nu, \quad 
Y+Y\to \{YY\} + \nu + \bar \nu, 
\end{eqnarray}
where $\{YY\}$ stand for a hyperonic Cooper pair. The PBF processes
were included also for nucleonic $S$- and $P$-wave superfluids. For
the $S$-wave $\Lambda$ condensate, we use the rate derived in Paper
I. For the $P$-wave hyperonic condensate we adopt the emission rate
derived for neutrons \citep{Leinson2017} with appropriate changes in
the condensate parameter and weak charges.

Our strategy in studying the cooling of hypernuclear compact stars is
to vary the masses of the stars in the range between 1 to
$2M_{\odot}$. As pointed out in Paper I, there is a mass hierarchy in
{ the} cooling behavior, with higher mass stars cooling faster than
low mass stars.  Another aspect of our study is to vary the input
pairing gaps in order to quantify their effect on cooling behavior.
We input three nucleonic gaps: the neutron $^1S_0$, neutron
$^3P_2$-$^3F_2$, and proton $^1S_0$ gap. These form a {\it triplet of
  nucleonic pairing gaps.} Our strategy will be to use alternate gap
inputs in such a triplet.  The first triplet is defined as SFB-0-CCDK,
where the acronyms SFB and CCDK refer to gaps of \cite{SFB_2003} and
\cite{Chen_NPA1993}, and { the number} 0 stands for zero pairing gap.
The second triplet of pairing gaps is DingS-DingP-CCDK, where DingS
and DingP refer to the $^1S_0$ and $^3P_2$-$^3F_2$ gaps obtained by
\cite{Ding_PRC_2016}.  The hyperonic $\Lambda$ $^1S_0$ pairing is
based on our BCS computation with background matter properties
specified by the corresponding EoS. It is added to each triplet of
nucleonic gaps. Contrary to what we have done in Paper I, here we
disregard pairing of $\Xi$ hyperons. The rationale is to not mask the
effect of the high-density pairing of protons and
$\Lambda$'s. $\Sigma^-$-pairing is also disregarded.  Simulations were
conducted with and without $P$-wave pairing for $\Lambda$'s and
protons in the high-density regime in order to quantify their effect;
their values were obtained via the scaling procedure described
above. In the case of SFB-0-CCDK/DingS-DingP-CCDK, $P$-wave pairing
for $\Lambda$'s and protons is calculated based on neutron $P$-wave
pairing of \cite{Baldo_PRC58}/\cite{Ding_PRC_2016}.

\begin{figure}
\begin{center}
\includegraphics[angle=0,width=0.99\columnwidth]{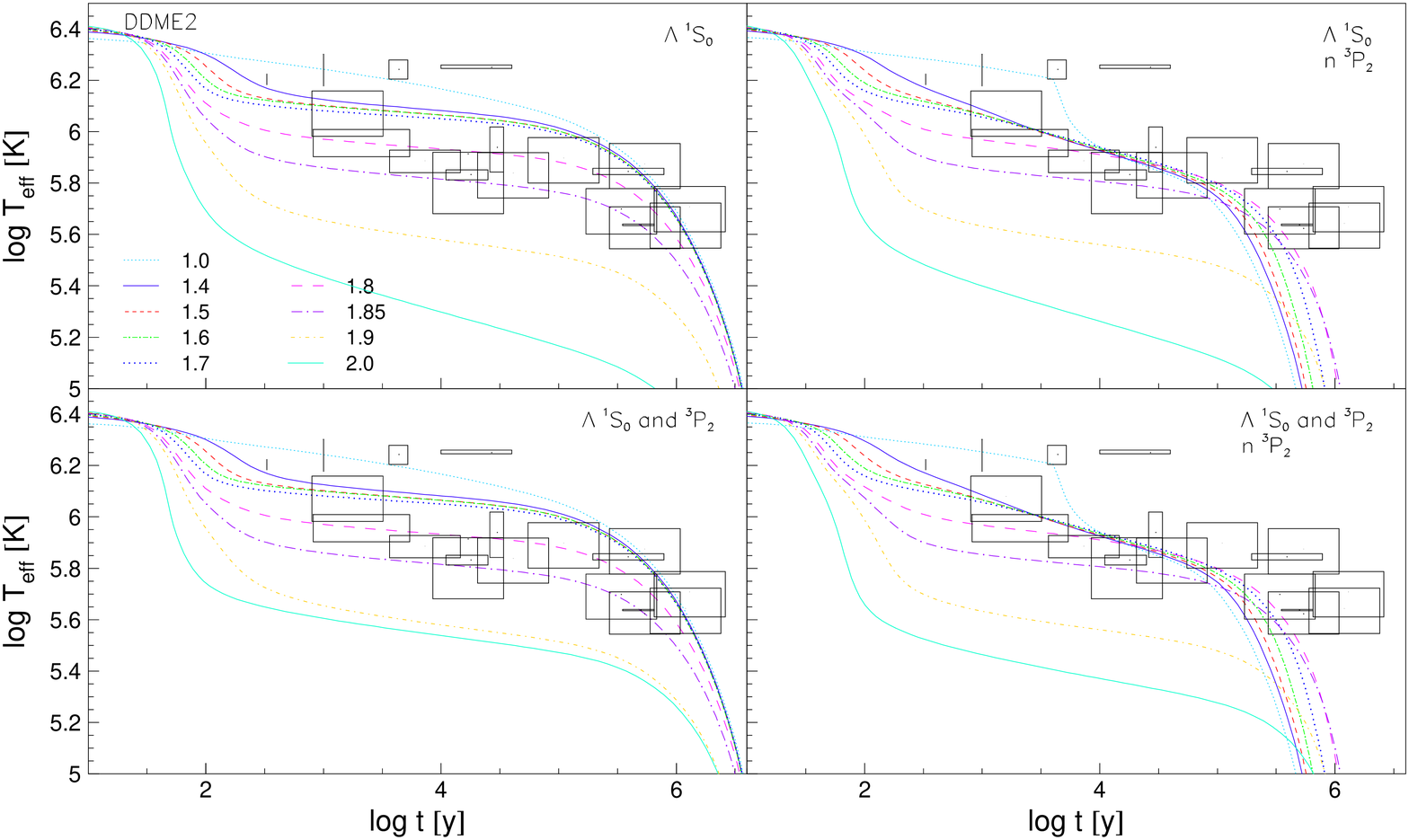}
\includegraphics[angle=0, width=0.99\columnwidth]{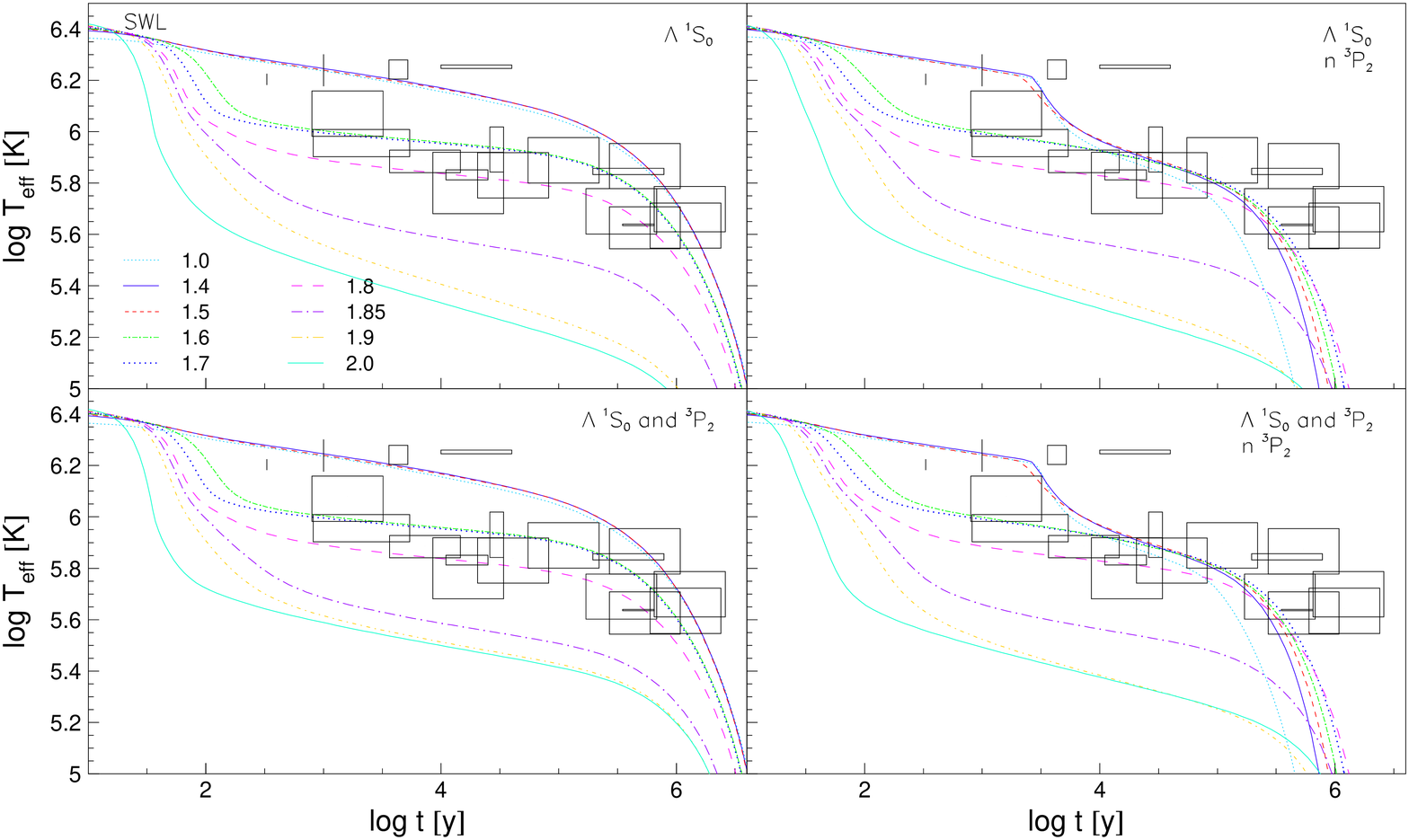}
\end{center}
\caption{Cooling models based on the DDME2 EoS (upper two rows) and
  SWL EoS (lower two rows) for NS masses within the range
  $1\le M/M_{\odot} \le 2$.  The left panels use for the triplet of
  neutron $^1S_0$ and $^3P_2$-$^3F_2$ and proton $^1S_0$ gaps { of/associated with}
  SFB-0-CCDK,  {where SFB refers to \protect\cite{SFB_2003}
  and CCDK to \protect\cite{Chen_NPA1993}, and 0 indicates that the neutron $P$-wave
  pairing gap is zero.} The right panels use the DingS-DingP-CCDK
  triplet. The $\Lambda$ $^1S_0$ pairing gaps are based on our BCS
  computations. The $\Lambda$ and proton $P$-wave gaps are obtained
  from the scaling procedure described in the text using for the
  neutron {$P$}-wave the results of {\protect\cite{Baldo_PRC58}} (left column) and,
  respectively, {\protect\cite{Ding_PRC_2016}} (right column).}
\label{fig:Teff_DDME2}
\end{figure}

\begin{figure}
\begin{center}
\includegraphics[angle=0,width=0.99\columnwidth]{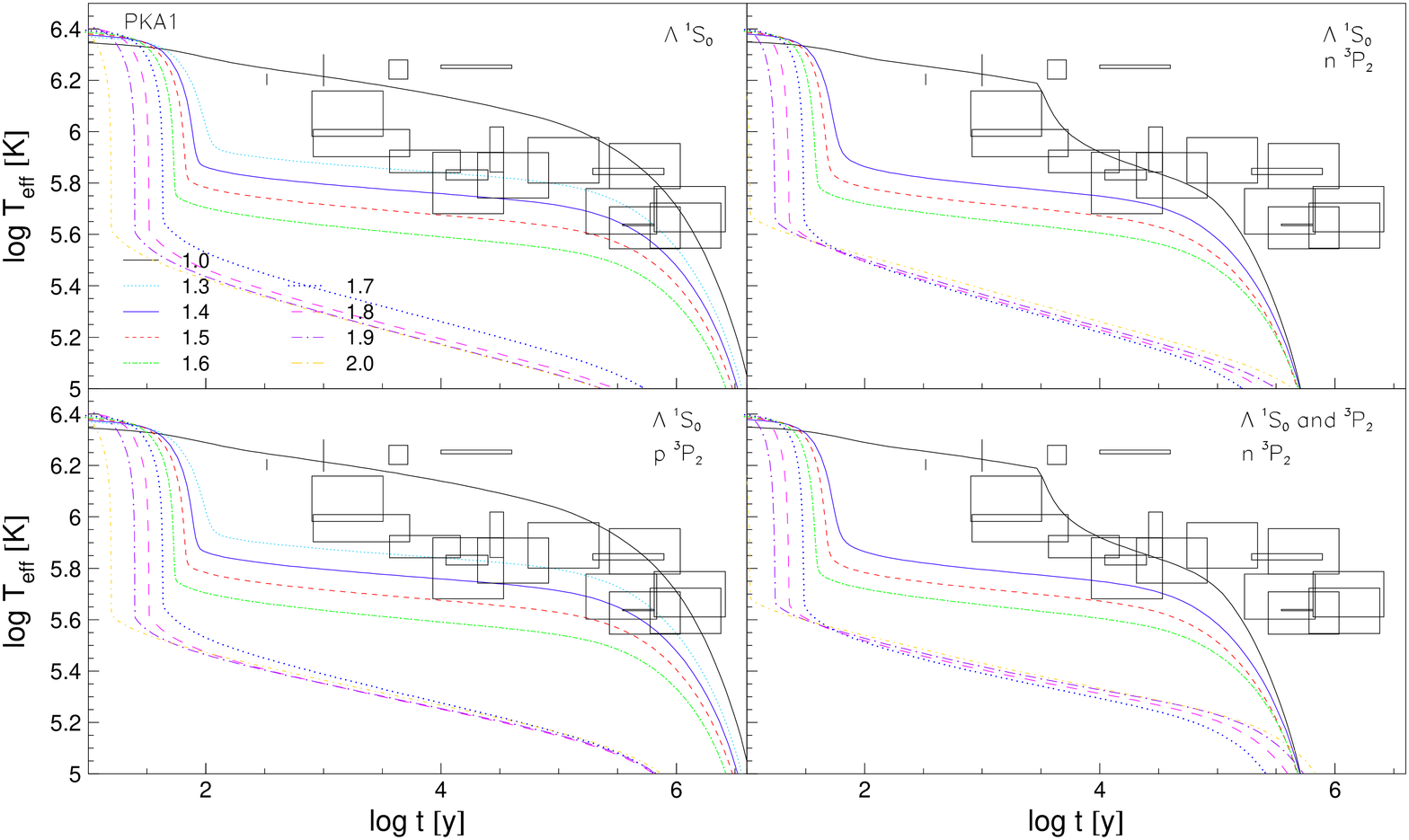}
\includegraphics[angle=0, width=0.99\columnwidth]{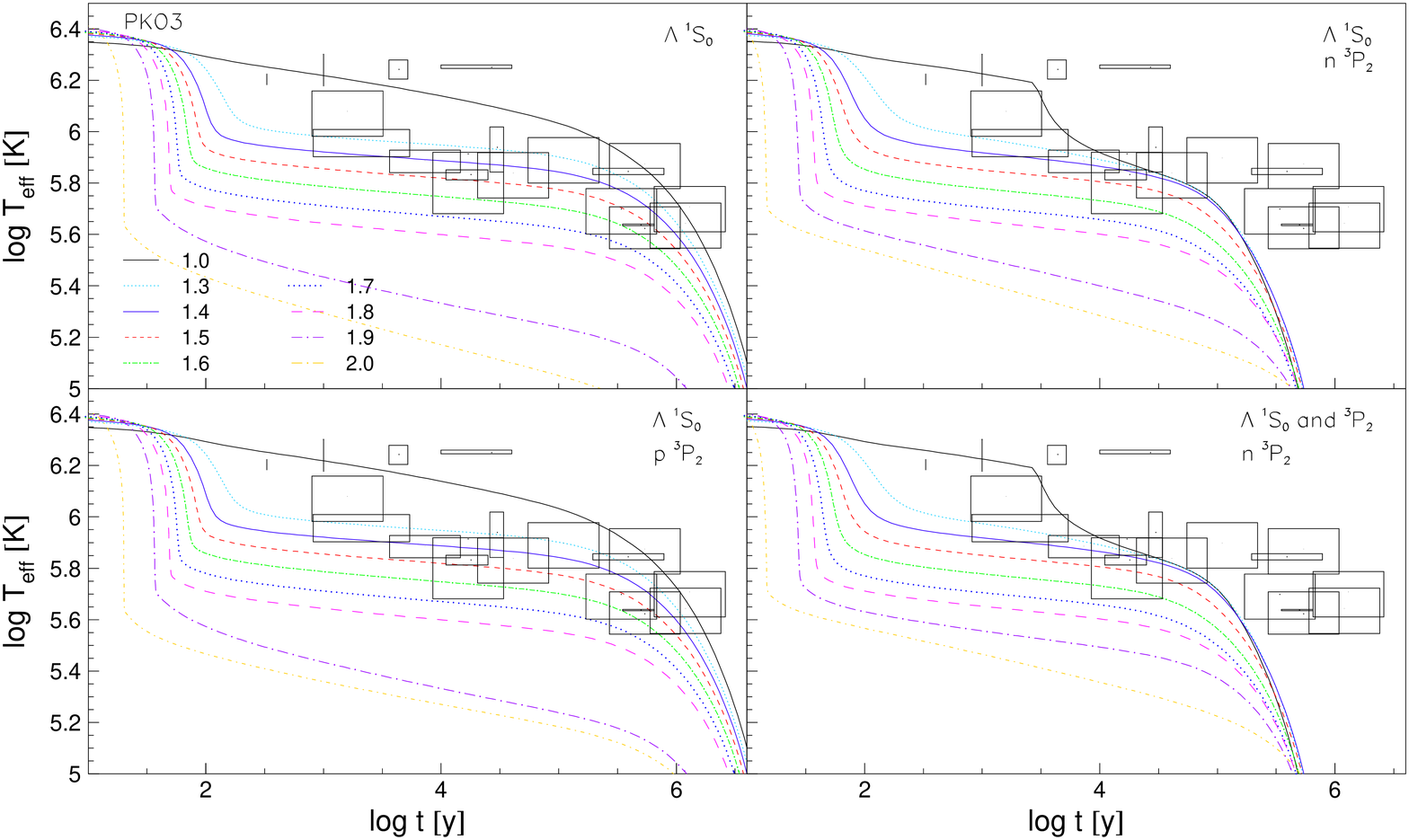}
\end{center}
\caption{Cooling models based on the PKA1 EoS (upper two rows) and
  PKO3 EoS (lower two rows) for NS masses within the range $1\le
  M/M_{\odot} \le 2$.  The left panels use for the triplet of neutron
  $^1S_0$ and $^3P_2$-$^3F_2$ and proton $^1S_0$ gaps  
  { of/associated with} SFB-0-CCDK, where the acronyms SFB refers to \protect\cite{SFB_2003}
  and CCDK to \protect\cite{Chen_NPA1993}, and 0 stands for zero
  pairing gap.  The right panels use the following triplet of pairing
  gaps: DingS-DingP-CCDK, where DingS and DingP refer to the $^1S_0$
  and $^3P_2$-$^3F_2$ gaps obtained by \protect\cite{Ding_PRC_2016}.
  The $\Lambda$ $^1S_0$ pairing gap are based on our BCS computations.
  The $\Lambda$ and proton $P$-wave gaps are obtained from the scaling
  procedure described in the text using for the neutron {$P$}-wave the
  results of {\protect\cite{Baldo_PRC58}} (left column) and, respectively,
  {\protect\cite{Ding_PRC_2016}} (right column).}
\label{fig:Teff_PKA1}
\end{figure}

Our survey of cooling models shows that they can be separated, broadly
speaking, into two groups. The first group which is characterized by
low-$L$ values are the Hartree models DDME2 and SWL. The second group
associated with large-$L$ values comprises the GM1A model and all the
HF models. As shown above, these two groups are characterized by high
and low nucleonic dUrca thresholds, respectively. Below we discuss
them separately and display for the second group only the results
obtained for the PKA1 and PKO3 EoS.  Figures~\ref{fig:Teff_DDME2} and
~\ref{fig:Teff_PKA1} show the cooling tracks for different mass models
for these two groups of EoS. For each group, the various panels include
or exclude the high-density $P$-wave pairing.

Before turning to the discussion of the features specific to different
EoS it is useful to point out the common features seen in the cooling
simulations. (a) There is a mass hierarchy with respect to the speed
of cooling - the heavier stars cool faster than the light ones.  (b)
Neutrino emission due to the PBF processes in a neutron $P$-wave
superfluid in the core leads to accelerated cooling after $t \ge 10^3$
yr. This acceleration is substantial even for the smallest gaps
adopted here (DingP).  As a consequence, the cooling tracks undershoot
the observational value of the temperature of XMMU J173203.3-344518.
Since it is expected that heating, accretion or magnetic fields do not
play a role for this object, its high temperature is inconsistent with
the existence of a neutron superfluid in the star's core. This
motivates the choice of a zero $P$-wave gap in our study. (c) Once
dUrca processes are allowed the cooling rate accelerates strongly; the
temperatures of the models are then determined by the interplay of
several factors: (i) existence of non-superfluid shells within the
stars; (ii) effectiveness of suppression of { the} dUrca process by
pairing gaps. The strength of the suppression depends on the
  magnitude and distribution of the gaps in the star.

  Let us now turn to a more detailed discussion of the cooling tracks
  in Figure~\ref{fig:Teff_DDME2} for stellar models based on the DDME2
  and SWL EoS. Table \ref{tab:DDME2} provides additional information
  on the reactions which are dominant in a star with a given mass. It
  also shows the effectiveness of high-density $P$-wave pairing of
  protons and $\Lambda$'s in slowing down the cooling.  For DDME2 EoS
  it is seen that, with the exception of the lightest ($1M_{\odot}$)
  star, all stars cool through the hyperonic dUrca processes and the
  PBF process in neutron $S$- and $P$-wave superfluids. (We recall
  that for the DDME2 EoS the proton fraction is below the nucleonic
  dUrca threshold { so that the dUrca process} does not occur in any
  of these mass models.)  Our calculations allow us to extract the
  density ranges of various processes: $(\Lambda, p)$ channel is
  active in the density range $0.34 \leq n_B \leq 1.02$ fm$^{-3}$
  ($M/M_{\odot} \geq 1.38$); the $(\Xi^-,\Lambda)$ channel is active
  in the range $0.37 \leq n_B \leq 0.98$ fm$^{-3}$
  ($ M/M_{\odot} \geq 1.54$); the $(\Sigma^-,\Lambda)$ channel is
  active in the range $0.39 \leq n_B \leq 0.60$ fm$^{-3}$
  ($1.60 \leq M/M_{\odot} \leq 2$).  Note that here and in all models
  that will be discussed below the $(\Sigma^-, n)$ channel is
  energetically forbidden.  This is not surprising given the large
  difference in their abundances.

  We observe that the mass range $1\le M/M_{\odot}\le 1.85$ covers the
  observed range of temperature well, showing a clear mass hierarchy
  of the cooling behavior of NSs in the neutrino cooling
  era $ t\le 10^5$ yr. [Note that this hierarchy is inverted at the
  very early ($t\le 10$ yr) stages of thermal evolution, which is
  however observationally insignificant].

  Firstly, we note that switching on and off the neutron $P$-wave
  pairing has an important effect on the cooling of
  low-to-intermediate mass stars as this pairing significantly reduces
  the heat capacity of the core and induces one of the dominant
  cooling processes via PBF.  For the late cooling era $t\ge 10^5$~yr,
  we find that the cooling tracks computed for zero neutron $P$-wave
  pairing gaps are in better agreement with the data than the cooling
  tracks computed for non-zero gaps. [Note that the youngest stars
  such as the compact central objects (CCO) in Cass A with an
  estimated age of $t \sim 330 $~yr or XMMU J173203.3-344518 with
  $t \sim 2.7 \times 10^3$~yr may have light elements in their
  atmospheres. Therefore, they  may have higher temperatures than
  predicted by our models.]  As can be seen from Table
  \ref{tab:DDME2}{,} the hyperonic dUrca processes operate  in all
  $M/M_{\odot} \geq 1.38$ stars.  Stars with
  $1.40 \leq M/M_{\odot} \leq 1.6$ cool down via $(\Lambda, p)$, 
    which is the only active dUrca process in their interiors.  In
    contrast to this, several reactions are triggered in stars with
    masses of $M/M_{\odot} \geq 1.7$, with
  $(\Sigma^-,\Lambda)$ being the dominant one.  Nevertheless, the
  cooling of stars with $M/M_{\odot} \leq 1.85$ is not fast because of
  the presence of a  proton and/or $\Lambda$ superfluidity.

  Increasing the mass of the star increases its central density. For a
  sufficiently massive star, its central density can exceed the density
  at which the $\Lambda$ $S$-wave gap disappears.  In this case the
  stars cool fast by the $(\Sigma^-,\Lambda)$ dUrca processes because
  at high densities $\Lambda$'s are unpaired.  However, if one allows
  for a higher $P$-wave pairing of $\Lambda$'s, it slows down the
  cooling of the $M/M_{\odot}=2$ model, as seen
  Fig.~\ref{fig:Teff_DDME2}.  The $\Lambda$ $P$-wave pairing scaled
  from neutron $P$-wave pairing results of \cite{Ding_PRC_2016} (right
  panel) is less effective than the one obtained from that of
  \cite{Baldo_PRC58} (left panel).  The reason is that the latter
  spans a wider range of densities and its maximum is located at a
  larger value (see Fig. \ref{fig:pwave_gaps}).

  Next, we consider the cooling tracks computed for the SWL EoS. They
  are shown in the lower part of Fig.~\ref{fig:Teff_DDME2}.  This case
  is similar to DDME2 considered above, as the nucleonic dUrca process
  does not occur in stars with masses $M/M_{\odot} \leq 2$.  From our
  models the density ranges of hyperonic dUrca processes are as
  follows: $(\Lambda,p)$ channel operates in the density range
  $0.41 \leq n_B \leq 0.94$ fm$^{-3}$ ($1.5 \leq M/M_{\odot} \leq 2$),
  $(\Xi^-,\Lambda)$ operates in the range $0.45 \leq n_B \leq 0.94$
  fm$^{-3}$ ($1.66 \leq M/M_{\odot} \leq 2$).  As was the case for the
  DDME2 EoS keeping or switching off the $P$-wave pairing of neutrons
  affects the cooling of low-to-intermediate mass stars due to the
  presence or absence of the corresponding PBF process. It can be
  seen that the zero gap case reproduces the measured surface
  temperatures of mature $t\ge 10^5$ NSs. Because
  hyperonic dUrca processes are dominant in stars with masses
  $M/M_{\odot} \geq 1.6$ (see Table~\ref{tab:DDME2}) the cooling
  tracks of stars with masses below this value are practically
  indistinguishable from each other and produce the highest
  temperatures. Once the hyperonic dUrca processes set in there is a
  mild drop in the cooling tracks of stars of masses 
  $M/M_{\odot} \leq 1.8$, produced by the balance between  the large
    neutrino emissivities of these hyperonic dUrca processes and 
    the suppression via hyperon and proton $S$-wave pairing. The
  onset of  the reaction $(\Xi^-,\Lambda)$, which is only partially
  suppressed, accelerates the cooling of  the most massive stars, 
    leading to a significant drop in  their temperature.  $P$-wave
  pairing of $\Lambda$'s results in higher temperatures of stars with
  $M/M_{\odot} \geq 1.9$, for it diminishes the emissivity of both
  $(\Xi^-,\Lambda)$ and $(\Lambda,p)$.  Contrary to what we have seen
  for the case DDME2 EoS based models, for the SWL EoS based models
  the $P$-wave pairing of protons plays a role too when the proton
  gap is large enough, as it happens when the gap is obtained from scaling 
  the neutron $P$-wave pairing gap of \cite{Baldo_PRC58}.
  This case is not illustrated in Fig. \ref{fig:Teff_DDME2}
  as it is similar to the case of $\Lambda$-P wave pairing. 

\onecolumn
\twocolumn
\begin{table*}
    \caption{List of reactions which dominate the cooling evolution
      (columns 2 and 5) and efficiency of high density pairing of
      protons and $\Lambda$s (indicated by $-$ (zero efficiency) or
      $\surd$ (non-zero efficiency) in columns 3, 4, 6, 7) in NS
      models of various masses built upon EoS which do not allow for
      nucleonic dUrca, DDME2 (upper part) and SWL (lower part), for
      any of the considered masses.  Columns 2-4 (5-7) correspond to
      the set SFB-0-CCDK (DingS-DingP-CCDK) for neutron $^1S_0$ { and}
       $^3P_2$-$^3F_2$ and proton $^1S_0$ gaps, see text.  The
      $\Lambda$ $^1S_0$ gaps are based on our BCS computations
      assuming { a} background given by the corresponding EoS.  The proton
      and $\Lambda$ $^3P_2$ gaps are obtained by scaling the neutron
      $P$-wave gap by \protect\cite{Baldo_PRC58} (columns 3-4) or,
      alternatively, by \protect\cite{Ding_PRC_2016} (columns 6-7).
      The abbreviations are as follows: PBF { stands for} Cooper pair
      breaking-formation processes; $n(^3P_2)$ { } neutron
      $P$-wave pairing and similarly for other baryons and channels.
      Baryons in brackets, such as $(\Lambda, p)$ stand for direct
      Urca process involving indicated baryons [see
        Eqs.~\eqref{eq:UrcaLambda}-\eqref{eq:Xizero}].  N
      bremstr. stands for nucleonic bremsstrahlung.}
    \label{tab:DDME2}
    \begin{tabular}{c|c|c|c|c|c|c|c}
      $M/M_{\odot}$ & dominant reactions & $p(^3P_2)$ & $\Lambda(^3P_2)$ & dominant reactions & $p(^3P_2)$ & $\Lambda(^3P_2)$ \\
      \hline
      &   \multicolumn{3}{c|}{no $n(^3P_2)$}  & \multicolumn{3}{c|}{$n(^3P_2)$} \\
      \hline 
    DDME2 &   &  &  &  &  &  \\
\hline
      1.0 & PBF\_n$^1S_0$, N bremstr. & $-$ & $-$ & PBF\_n$^3P_2$, N bremstr. & $-$ & $-$\\
      1.4 & $(\Lambda,p)$, PBF\_n$^1S_0$ & $-$ & $-$ & $(\Lambda,p)$, PBF\_n$^3P_2$ & $-$ & $-$\\
      1.5 & $(\Lambda,p)$, PBF\_n$^1S_0$ & $-$ & $-$& $(\Lambda,p)$, PBF\_n$^3P_2$ & $-$ & $-$ \\
      1.6 & $(\Lambda,p)$, PBF\_n$^1S_0$ & $-$ & $-$& $(\Lambda,p)$, PBF\_n$^3P_2$ & $-$ & $-$\\
      1.7 & $(\Lambda,p)$, $(\Sigma^-,\Lambda)$ ,PBF\_n$^1S_0$ & $-$ & $-$& $(\Lambda,p)$, PBF\_n$^3P_2$ & $-$ & $-$\\
      1.8 & $(\Sigma^-,\Lambda)$, PBF\_n$^1S_0$ & $-$ & $-$& $(\Sigma^-,\Lambda)$ ,$(\Lambda,p)$, PBF\_n$^3P_2$ & $-$ & $-$\\
      1.9 & $(\Sigma^-,\Lambda)$  & $-$ & $-$ & $(\Sigma^-,\Lambda)$, $(\Lambda,p)$ & $-$ & $-$\\
      2.0 & $(\Sigma^-,\Lambda)$  & $-$ & $\surd$ & $(\Sigma^-,\Lambda)$, $(\Lambda,p)$ & $-$ &  $\surd$\\
      \hline 
     SWL &   &  &  &  &  &  \\
\hline
      1.3 & PBF\_n$^1S_0$, N bremstr. & $-$ & $-$ & PBF\_n$^3P_2$, N bremstr. & $-$ & $-$ \\
      1.4 & PBF\_n$^1S_0$, N bremstr. & $-$ & $-$ & PBF\_n$^3P_2$, N bremstr. & $-$ & $-$\\
      1.5 & PBF\_n$^1S_0$, N bremstr. & $-$ & $-$ & PBF\_n$^3P_2$, N bremstr. & $-$ & $-$\\
      1.6 & $(\Lambda,p)$, PBF\_n$^1S_0$ & $-$ & $ -$ & $(\Lambda,p)$, PBF\_n$^3P_2$  & $-$ & $-$\\
      1.7 & $(\Lambda,p)$, PBF\_n$^1S_0$ & $-$ & $-$ & $(\Lambda,p)$, PBF\_n$^3P_2$  & $-$ & $-$\\
      1.8 & $(\Lambda,p)$, $(\Xi^-,\Lambda$) & $-$ & $-$ & $(\Lambda,p)$, ($\Xi^-,\Lambda)$, PBF\_n$^3P_2$  & $-$ & $-$\\
      1.9 & $(\Lambda,p)$, $(\Xi^-,\Lambda$) & $\surd$ & $\surd$ & $(\Lambda,p)$, ($\Xi^-,\Lambda)$  & $-$ & $\surd$\\
      2.0 & $(\Lambda,p)$, $(\Xi^-,\Lambda$) & $\surd$ & $\surd$ & $(\Lambda,p)$, ($\Xi^-,\Lambda)$  & $-$ & $\surd$\\
 \hline
    \end{tabular}
\end{table*}
\onecolumn
\twocolumn

\begin{table*}
  \begin{center}
    \caption{Same as in Table \ref{tab:DDME2} but for the PKA1 and PKO3 EoS. 
    }
    \label{tab:PKO3}
    \begin{tabular}{c|c|c|c|c|c|c|c}
\hline
      $M/M_{\odot}$ & dominant reactions & $p(^3P_2)$ & $\Lambda(^3P_2)$ & dominant reactions & $p(^3P_2)$ & $\Lambda(^3P_2)$ \\
      \hline
      &   \multicolumn{3}{c|}{no $n(^3P_2)$}  &
                                                \multicolumn{3}{c|}{$n(^3P_2)$} \\
     \hline 
    PKA1 &   &  &  &  &  &  \\
\hline
      1.0 & PBF\_n$^1S_0$, N bremstr. & $-$ & $-$ & PBF\_n$^1S_0$, PBF\_n$^3P_2$, N bremstr. & $-$ & $-$ \\
      1.4 & $(n,p)$  & $-$ & $-$ & $(n,p)$  & $-$ & $-$\\
      1.5 & $(n,p)$  & $-$ & $-$ & $(n,p)$  & $-$ & $-$\\
      1.6 & $(n,p)$  & $-$ & $-$ & $(n,p)$, $(\Lambda,p)$  & $-$ & $-$\\
      1.7 &  $(n,p)$, $(\Sigma^-,\Lambda)$  & $\surd$  & $\surd$ & $(\Sigma^-,\Lambda)$  & $-$  & $\surd$ \\
      1.8 & $(n,p)$ & $\surd$ & $-$ & $(\Sigma^-,\Lambda)$  & $-$  & $\surd$\\
      1.9 & $(n,p)$ & $\surd$ & $-$ & $(\Sigma^-,\Lambda)$  & $-$  & $\surd$\\
      2.0 &$(n,p)$ & $\surd$ & $-$ & $(\Lambda,p)$, $(\Sigma^-,\Lambda)$, $(\Xi^-, \Lambda)$  & $-$  & $\surd$\\
   \hline
    PKO3 &   &  &  &  &  &  \\
\hline
      1.0 & PBF\_n$^1S_0$, N bremstr. & $-$ & $-$ & PBF\_n$^3P_2$, N bremstr. & $-$ & $-$ \\
      1.4 & $(n,p)$  & $-$ & $-$ & $(n,p)$, PBF\_n$^3P_2$ & $-$ & $-$ \\
      1.5 & $(n,p)$  & $-$ & $-$ & $(n,p)$, PBF\_n$^3P_2$ & $-$ & $-$ \\
      1.6 & $(n,p)$  & $-$ & $-$ & $(n,p)$ & $-$ & $-$ \\
      1.7 & $(n,p)$  & $-$ & $-$ & $(n,p)$ & $-$ & $-$ \\
      1.8 & $(n,p)$  & $-$ & $-$ & $(n,p)$, $(\Lambda,p)$ & $-$ & $-$\\
      1.9 & $(n,p)$  & $-$ & $-$ & $(n,p)$, $(\Lambda,p)$ & $-$ & $\surd$\\
      2.0 & $(n,p)$  &$\surd$  & $-$&  $(\Lambda,p)$, $(\Xi^-,\Lambda)$, & $-$ & $\surd$ \\
   \hline
  \end{tabular}
  \end{center}
\end{table*}
The cooling tracks for the PKA1 and PKO3 EoS are shown in
Fig.~\ref{fig:Teff_PKA1}. Compared to the 
DDME2 and SWL based models discussed above, 
the new qualitative feature is the low threshold for
the nucleonic direct Urca process.

We extract from our PKA1 EoS a density threshold of $n_B
\geq 0.25$ fm$^{-3}$ above which the
$(n,p)$
reaction is effective. This density corresponds to $M/M_{\odot}
\geq
0.98$, which means that all models considered allow for nucleonic
dUrca.  Additional contributions come from the hyperonic dUrca
processes, where $(\Lambda,p)$
is effective over the density range $0.32
\leq n_B \leq 0.95$ fm$^{-3}$ ($1.44 \leq M/M_{\odot} \leq
1.99$), $(\Sigma^-,\Lambda)$ is effective over the range $0.40 \leq
n_B \leq 0.47$ fm$^{-3}$ ($1.59 \leq M/M_{\odot} \leq
1.67$), and $(\Xi^-,\Lambda)$ acts over the range $0.45 \leq n_B \leq
0.95$ fm$^{-3}$ ($1.64 \leq M/M_{\odot} \leq
1.99$).  The cooling tracks, illustrated in the top panels of
Fig. ~\ref{fig:Teff_PKA1}, fall into three classes.  The first class
is represented by the lightest star.  With a core that accommodates
nucleonic dUrca over a volume of insignificant size, the
$M/M_{\odot}=1$
model manifests slow cooling, via neutrino emission from nucleon
bremsstrahlung and PBF from neutron pairing in {{$S$-
    and $P$}}-channels.
The second class of models, with $1.3
\leq M/M_{\odot} \leq
1.6$, cool down exclusively or predominantly by the nucleonic dUrca
process.  As at least one of the involved species, { i.e.}  protons,
$\Lambda$s
or - { as in the case for the panels on the right} - neutrons, is
paired{,} the cooling is moderate and the cooling tracks { cover} a
domain in the $t$-$T_{\rm
  eff}$ plane where { observed} data exist or are not far from them.
The third class is represented by models with $M/M_{\odot}
\geq
1.7$.  Their fast cooling is due to unpaired baryons, protons (left
panels of Fig. ~\ref{fig:Teff_PKA1}) and $\Lambda$s
(right panels of Fig. ~\ref{fig:Teff_PKA1}).  As already discussed in
relation with the DDME2- and SWL-models, even a small neutron
$P$-wave
pairing leads to accelerated cooling for $t
\gtrsim
10^3$ yr, irrespective of the star's mass.  According to Table
\ref{tab:PKO3}, even a tiny neutron $P$-wave
pairing gap (or, equivalently $T_c$)
as the one of \cite{Ding_PRC_2016} is sufficient to switch the
dominant dUrca from nucleonic (left panels, no neutron $P$-pairing)
to hyperonic (right panels, with neutron $P$-pairing).

Switching on and off the $P$-wave superfluidity in the proton and
$\Lambda$ components does not change the general behavior of cooling
tracks qualitatively. It is seen that most massive stars with $P$-wave
superfluidity have higher temperatures compared to their counterparts
without such pairing. Depending on whether or not neutron $P$-wave
pairing occurs, the cooling of these stars is slowed down by proton-
or, alternatively, $\Lambda$-$P$-wave pairing.

The models based on the PKO3 EoS, shown in the bottom panels of
Fig. ~\ref{fig:Teff_PKA1}, are qualitatively similar to those based on
the PKA1 EoS.  The reason for this lies in the relatively low
nucleonic dUrca threshold, $n_B \geq 0.28$ fm$^{-3}$, which makes this
channel effective in models with $M/M_{\odot} \geq 1.22$.  The
$(\Lambda,p)$ and $(\Xi^-,\Lambda)$ hyperonic dUrca processes act over
baryonic density ranges of $0.33 \leq n_B \leq 0.86$ fm$^{-3}$ and
$0.54 \leq n_B \leq 0.83$ fm$^{-3}$, respectively, which correspond to
NS masses domains of $1.56 \leq M/M_{\odot} \leq 1.99$ and
$1.89 \leq M/M_{\odot} \leq 1.99$.

As for PKA1 EoS based models, the nucleonic dUrca in stars with
$M/M_{\odot} \geq 1.8$ is suppressed by the neutron $P$-wave pairing;
as a consequence, the hyperonic dUrca, in this case $(\Lambda,p)$, is
as efficient as the nucleonic dUrca.  As illustrated in
Fig. \ref{fig:kF_NSM}, PKO3 EoS leads to lower proton- and
$\Lambda$-abundances than PKA1 EoS.  As a consequence, the
$^1S_0$-pairing of protons and $\Lambda$'s computed for stellar models
based on the PKO3 EoS extends over larger core regions than is the
case for models based on the PKA1 EoS.  They are additionally
characterized by higher critical temperatures, as can be seen in
Figs.~\ref{fig:Tcp} and \ref{fig:TcL}.  These facts explain the
quantitative differences between the cooling curves associated with
the two EoS.  The separation between slow-, intermediate- and
fast-cooling tracks is less pronounced for PKO3 than for PKA1 based
EoS.  We also note that for a given stellar mass and fixed set of
pairing-gaps, the effective temperatures are higher.  The lower
densities of protons and $\Lambda$'s produced by PKO3 render the
high-density pairing of these particles less efficient. Indeed, the
drop in the temperature becomes smaller only for models with
$M/M_{\odot}=1.9$ and 2 when $\Lambda$ $P$-wave pairing is
additionally taken into account; this occurs only for $M/M_{\odot}= 2$
models if proton $P$-wave pairing is included in addition.

\section{Conclusions}
\label{sec:conclusions}

This work extends the previous study of \cite{Raduta2018MNRAS} to the
cooling of hypernuclear stars by including models for the EoS based on
the Hartree-Fock treatment of the hypernuclear matter.  We conjecture
on the existence of new channels of $P$-wave pairing in the proton and
$\Lambda$ components of our models and explore their effect on the
cooling behavior of NSs using simple scaling relations for the pairing
gaps. The EoS models can be classified into two groups: the first
group, which includes the DDME2 and SWL models, has a low-$L$ value
and does not allow for nucleonic dUrca or allows for it only in very
massive stars.  The second group consists of GM1A and all HF models,
which have a high $L$ value but a rather low nucleonic dUrca
threshold.

The grouping of the models in two classes reflects the
  uncertainties in predictions of the current density functionals of
  the $L$-value, which is not well constrained.  The variations in the
  $L$ value predicted by different models can be traced back to the
  fit parameters that are used to construct any particular density
  functional, in particular, the relative weight of observables that
  belong to iso-vector and iso-scalar sectors.  In low-$L$ density
  functionals (e.g. the DDME2 functional) the iso-vector probes have a
  significant weight in the data to which the parameters are tuned.
  On the contrary, the family of Hartree-Fock PK-models, which predict
  large $L$-values, has been designed using fits to experimental data
  of binding energies of magic nuclei, spin-orbit splittings of
  neutron and proton 1p states of $^{16}{\rm O}$, shell gaps at $Z=58$
  and 64 and properties of symmetric saturated matter.  

\begin{itemize}
\item \underline{Mass hierarchies and observational data.}  We find in
  general that all the models follow a mass-hierarchy.  If one
  arranges the stars from the lightest to the heaviest ones, the
  former cool always slower than the latter in the neutrino cooling
  era $t< 10^5$~yr.  Mass hierarchies have been observed for a
    long time, in particular in the studies where some fast cooling
    agent has been included; for example, this is the case when stars
    contain a quark
    core~\citep{Page2000,Grigorian2005,Blaschke2012,Hess2011PhRvD,Sedrakian2013}
    or cool by direct Urca
    process~\citep{Schaab1997AA,Taranto2016}. Such a hierarchy has also been
    also obtained in the studies based on the medium-modified cooling
    scenario; for a recent discussion see
    \citep{Grigorian_NPA_2018}.  The mass hierarchy may, however, be
  inverted at a very early stage of evolution, i.e., the heavier stars
  may cool slower than the lighter stars.  Similar inversion also is
  observed at very late stages, when the cooling is dominated by
  photon emission from the stellar surface (photon cooling era). For
  the DDME2 and SWL EoS based models, we find that cooling tracks
  smoothly cover the observationally relevant area in the
  $t$-$T_{\rm eff}$ diagram. Interestingly, we find that vanishing
  $P$-wave neutron superfluidity improves the agreement with the
  observational data for our hypernuclear models, a feature already
  observed in alternative models of cooling evolution of NS. The
  Hartree-Fock models develop nucleonic dUrca processes rather
  early. This results in a very fast cooling of these models which
  renders the agreement with the data difficult. Nevertheless, for the
  PKO3 model, we find that light to medium-mass stars have cooling
  tracks that pass through the observationally relevant region in the
  $t$-$T_{\rm eff}$ diagram.

\item \underline{High-density pairing.}  In this part of the work, we
  attempted to rectify the too rapid cooling of the most massive NS
  models of our study, which is linked to the closing of the $^1S_0$
  gaps for $\Lambda$'s and protons at high densities. We have proposed
  simple scaling arguments for obtaining the gaps of these particles
  from those computed for neutron matter. From this, we found that if
  $P$-wave pairing of protons and $\Lambda$'s is allowed, then even
  the most massive ($1.9\le M/M_{\odot}\le 2$) hypernuclear stars will
  undergo a significantly slower cooling evolution.

\end{itemize}

The diversity among presently considered hypernuclear EoS, which leads
to differences in the thermal evolution of NSs, is partly rooted in
the uncertainties in the nucleonic sector ($L$-value and, very
probably, higher order terms of the expansion of the symmetry energy
as a function of scalar density), the assumed spin-flavor symmetry of
the quark model which controls the coupling of hyperons to nucleons in
the vector meson sector, and the uncertainty in the values of the
hyperon-nucleon interaction potential which does the same in the
scalar meson sector.  An interesting possibility not studied so far is
a lower value of $\Sigma$ potential in the nuclear matter than assumed
here.  In that case, already relatively light mass stars will
experience fast cooling via $\Sigma^- \to \Lambda + l +\bar \nu_l$,
even if $\Lambda$-pairing is taken into account.  For DDME2 our
preliminary results show that models with $M/M_{\odot} \geq 1.5$
manifest strong disagreement with the data.  The situation might
nevertheless significantly improve if one additionally allows for
$\Sigma$-pairing. These scenarios are under consideration and will be
published elsewhere.  Despite of this diversity, our study elucidates
how various ingredients of microphysical input (parametrization of the
EoS, pairing gaps, etc.) affect the cooling evolution of neutron
stars. This establishes controlled correspondence between various
input parameters and the general behavior of cooling tracks of
hypernuclear stars, which can be used as a diagnostics of the internal
composition of NSs in the future. While a general picture of the
  cooling of hypernuclear star is emerging, the {\it systematic coverage}
  of the parameter space remains a task for the future.
  This would include the empirical parameters of nuclear matter at saturation, the
  range of pairing gaps predicted by the many-body theory, as well as
  an estimate of the effect of the statistical uncertainties that are
  inherent to the microphysics input on the cooling simulations.

\section*{Acknowledgments}
A. R. acknowledges the support provided by the European COST Actions
``NewCompStar'' (MP1304) and ``PHAROS'' (CA16214), through STSM grants
as well as the hospitality of the Frankfurt Institute for Advanced
Studies.
J.-J.~Li was supported by Alexander von Humboldt Foundation. 
A. S. is supported by the Deutsche Forschungsgemeinschaft (Grant
No. SE 1836/4-1).
F. W. is supported by the National Science Foundation
(USA) under Grant PHY-1714068.  
This research has made use of NASA's Astrophysics Data System.

\bibliographystyle{mnras}
\bibliography{hypercooling_full.bib}{}
\end{document}